\documentclass{article}

\usepackage{amsfonts}
\usepackage{amsmath}
\usepackage{amsthm}
\usepackage{amssymb}
\usepackage{graphicx}

\def\be{\begin{eqnarray}}
\def\ee{\end{eqnarray}}
\numberwithin{equation}{section}

\newcommand{\eq}[2]{\begin{equation}\label{#1}#2\end{equation}}

\begin{document}

\hfill ITEP/TH-70/08

\bigskip

\centerline{\Large
{Resultant as Determinant of Koszul Complex}}

\bigskip

\centerline{A.Anokhina, A.Morozov
and Sh.Shakirov\footnote{SashaAnokhina@yandex.ru;\
morozov@itep.ru;\ shakirov@itep.ru}}

\bigskip

\centerline{{\it ITEP, Moscow, Russia}}

\centerline{{\it MIPT, Dolgoprudny, Russia}}
%\title{Resultant as Determinant of Koszul Complex}
%\author{A.Anokhina, A.Morozov and Sh.Shakirov}%\date{}\maketitle

\bigskip

\bigskip

\begin{abstract}
A linear map between two vector spaces has a very
important characteristic: a determinant.
In modern theory two generalizations of linear maps
are intensively used: to linear complexes (the
nilpotent chains of linear maps) and to
non-linear mappings.
Accordingly, determinant of a linear map has two
generalizations: to determinants of complexes
and to resultants.
These quantities are in fact related:
resultant of a non-linear map is determinant of
the corresponding Koszul complex.
We give an elementary introduction into these
notions and interrelations, which will definitely
play a role in the future development of
theoretical physics.
\end{abstract}

\bigskip

\tableofcontents

\newpage

\section{Introduction}
The origin of matrices and determinants lies deep in ancient civilizations. Around the second century BC, the Babylonians studied linear equations and the Chineese used matrices to find solutions to simple systems of linear equations, in two or three unknowns \cite{HISTORY}. However, only in the late 17-th century these ideas became widely known and flourished with different applications. The
theory of \emph{linear} equations and transformations, now known as
\emph{linear algebra}, has completely proved its worth not only in
pure science, but also in applied problems and in engineering. For
example, Gauss himself applied linear algebra to determine the orbit of
the asteroid Pallas \cite{GAUSS}. Using observations of Pallas taken between 1803
and 1809, Gauss obtained and solved a system of six linear equations
in six unknowns.

Soon after that, it became clear to researchers (Cayley,
Sylvester and others \cite{CAYLEY,SYLV}) that linear algebra is just the
simplest case of more general framework -- of \emph{polynomial}
equations and transformations. They realized, that many parts of
linear algebra have natural generalizations to arbitrary
polynomials, not necessarily linear. Notably, determinant has such a
generalization. A linear system in two unknowns
$$ \left\{ \begin{array}{c}
f_{0} x + f_{1} y = 0 \\
\noalign{\medskip}g_{0} x + g_{1} y = 0 \\
\end{array} \right. $$
is solvable (has a non-trivial solution $x, y$) if and only if its determinant vanishes
\begin{center}
$f_{0} g_{1} - f_{1} g_{0} = 0$
\end{center}
and in complete analogy, a quadratic system
\[ \left\{ \begin{array}{c}
f_{0} x^2 + f_{1} x y + f_{2} y^2 = 0 \\
\noalign{\medskip}g_{0} x^2 + g_{1} x y + g_{2} y^2 = 0 \\
\end{array} \right. \]
is solvable if and only if certain expression, depending on coefficients, vanishes:
\begin{center}
$f_{0}^2 g_{2}^2 - f_{1} f_{2} g_{0} g_{1} + f_{1}^2 g_{0} g_{2} - 2 f_{0} f_{2} g_{0} g_{2} + f_{0} f_{2} g_{1}^2 - f_{0} f_{1} g_{1} g_{2} + f_{2}^2 g_{0}^2 = 0$
\end{center}
This expression -- a generalization of $2 \times 2$ determinant from linear to quadratic equations
-- is called a \emph{resultant}. It was Sylvester who first studied it \cite{SYLV} and found the analogues for cubic,
quartic and equations of higher degrees.
Resultants also exist for more than two variables:
in this case they provide non-linear geneneralizations of $n \times n$ determinant.
A system
\begin{equation}
\left\{ \begin{array}{c}
f_1(x_1,x_2,\ldots,x_n) = 0 \\
\noalign{\medskip} f_2(x_1,x_2,\ldots,x_n) = 0 \\
\noalign{\medskip} \ldots \\
\noalign{\medskip} f_n(x_1,x_2,\ldots,x_n) = 0 \\
\end{array} \right. \label{1}
\end{equation}
\smallskip\\
of $n$ homogeneous polynomials of degree $r$ in $n$ variables
$$ f_i \left( x_1, x_2, \ldots, x_n \right) = \sum\limits_{j_1, \ldots, j_r = 1}^{n} f_i^{j_1, j_2,...,j_r} x_{j_1} x_{j_2} ... x_{j_r} $$
is solvable, if and only if an expression called its resultant vanishes:
$$ R_{n|r}\{f_1, f_2, \ldots, f_n\} = 0 $$
For linear systems resultant $R_{n|1}$ is just a determinant,
the well-known and well-understood object of linear algebra.
Unfortunately, the knowledge and understanding of general resultants is not equally deep, yet.
Moreover, the very direction of generalization -- from linear algebra to arbitrary algebraic equations
-- has been almost completely forgotten after the 19-th century, and was revived only recently,
after profound book of Gelfand, Kapranov and Zelevinsky \cite{GKZ}.

For several reasons, we suggest to call this direction \emph{non-linear algebra} \cite{NOLINAL}. It is expected to have as many applications as its linear cousin, if not more. Resultants -- the main "special functions" of non-linear algebra -- are already used in physics to describe,
for example, multiparticle entanglement \cite{APP1}, singularities of Calabi-Yau three-folds \cite{APP2},
discrete dynamics \cite{APP3}. There are applications to modeling and robotics \cite{APP4}.
However, applications remain restricted, partially because calculation of resultants is not as easy as that
of determinants, especially when the number of variables is large.

\pagebreak

It is often convenient to speak about systems of equations in terms of maps. Linear algebra studies linear maps between two spaces of dimensions $m$ and $n$
$$ L_{1} \mathop{\rightarrow}^{f} L_{2} $$
which are given by linear polynomials:
$$ f: \left( \begin{array}{c} x_1 \\ x_2 \\ \ldots \\ x_n \end{array} \right) \mapsto \left( \begin{array}{c} f_{11} x_{1} + f_{12} x_{2} + \ldots + f_{1n} x_{n} \\ f_{21} x_{1} + f_{22} x_{2} + \ldots + f_{2n} x_{n} \\ \ldots \\ f_{m1} x_{1} + f_{n2} x_{2} + \ldots + f_{mn} x_{n} \end{array} \right) $$
Accordingly, non-linear algebra studies non-linear (polynomial) maps:
\begin{equation} f: \left( \begin{array}{c} x_1 \\ x_2 \\ \ldots \\ x_n \end{array} \right) \mapsto \left( \begin{array}{c} f_1(x_1,x_2,\ldots,x_n) \\ f_2(x_1,x_2,\ldots,x_n) \\ \ldots \\ f_m(x_1,x_2,\ldots,x_n) \end{array} \right) \label{2} \end{equation}
System (\ref{1}) corresponds to the particularly interesting case of $m = n$.
When the system (\ref{1}) is solvable, the corresponding map (\ref{2}) is degenerate, i.e, has a non-vanishing kernel.
As one can see, resultant defines a degeneracy condition for polynomial maps. Of course, it is very interesting to study further generalizations: to arbitrary non-linear mappings (not necessarily polynomial). For such generalizations and appropriate notion of resultant, see \cite{CONTOUR,RIEM}.

The second, essentially different, generalization of determinant was introduced by Cayley approximately at the same time \cite{CAYLEY}. Instead of a single linear map between two spaces
$$ L_{1} \mathop{\rightarrow}^{f} L_{2} $$
he considered a \emph{complex} -- sequence of two linear maps between three spaces
$$ L_{1} \mathop{\rightarrow}^{f_1} L_{2} \mathop{\rightarrow}^{f_2} L_{3} $$
subject to \emph{nilpotency} constraint $ f_{2} \circ f_{1} = 0 $. Generally, the number of spaces can be arbitrary
$$ L_{1} \mathop{\rightarrow}^{f_1} L_{2} \mathop{\rightarrow}^{f_2} L_{3} \mathop{\rightarrow}^{f_3} \ldots  $$
and the maps satisfy $ f_{i + 1} \circ f_{i} = 0 $ for all $i$.
Cayley has found a quantity, associated to complexes in the same way as determinant is associated to linear maps.
This quantity is now called \emph{determinant of a complex}. For example, consider a sequence of two maps between spaces of dimensions 1, 2 and 1:

$$ f_1: \ x \mapsto \left( \begin{array}{c} a_1 x \\ a_2 x \end{array} \right), \ \ \ f_2:\ \left( \begin{array}{c} y_1 \\ y_2 \end{array} \right) \mapsto b_1 y_1 + b_2 y_2 $$
\smallskip\\
Composition of these maps

$$ f_2 \circ f_1:\ x \mapsto ( a_1 b_1 + a_2 b_2 ) x $$
\smallskip\\
is requested to vanish. Cayley's determinant $\mbox{ DET } (f_1, f_2)$ can be represented in three different ways
$$ \mbox{ DET } (f_1, f_2) = \dfrac{a_2}{b_1} = - \dfrac{a_1}{b_2} = \dfrac{|b|}{|a|} $$
where $|a|$ is simply the Euclidean length of vector $a = (a_1,a_2)$. Equivalence of all three follows from the above condition $a_1 b_1 + a_2 b_2 = 0$. This example shows, that determinant of complex is not a polynomial, it is a rational function. It vanishes when $|b| = 0$ and is singular when $|a| = 0$. It may be not obvious at the moment, but we will see below, that such a quantity is indeed a natural generalization of determinant and has a number of important properties. In particular, determinant of a complex is (up to overall sign) invariant under permutations of basis vectors in the linear spaces.

It should be emphasized, that complexes are essentially objects of linear algebra and can be effectively treated using linear methods.
For this reason the theory of complexes and their determinants,
which is now called \emph{homological algebra}, is more developed and
attracts incomparably more attention, than resultant theory.
Homological algebra has a wide range of physical applications,
from topological field theories and invariants \cite{TOPOLOG} to
Faddeev-Popov's ghosts and Batalin-Vilkovisky approach to quantizing
particle Lagrangians and Lagrangians of string field theory \cite{BV}
and further to description of branes in terms of derived categories
\cite{bra}.

It has been known since Cayley, that resultants of non-linear maps and determinants of complexes are, in fact, related.
Namely, given a non-linear map one can construct a complex, called \emph{Koszul complex}, whose determinant is equal to resultant of the original map. Actually, this relation is one of a few reliable ways to calculate resultants, known at the moment. Several other ways to calculate resultants, not involving complexes, are described in \cite{LOGDET, CONTOUR}. The aim of this paper is to explain this relation in clear and practical terms, making the text accessible for beginners. It can be considered as pure pedagogical, since most of the results are quite classical and known to the experts (see, e.g. \cite{GKZ,EISENBUD,ALGO}).

\section{Resultants}

\subsection{Properties of resultants}
Thus it is interesting and important to calculate resultants. Before discussing the calculation methods, we formulate
the basic properties of resultants. As mentioned above, resultant $ R_{n|r}\{f_1, f_2, \ldots, f_n\} $ defines a degeneracy condition for polynomial maps of degree $r$ in $n$ variables
$$f: \left( \begin{array}{c} x_1 \\ x_2 \\ \ldots \\ x_n \end{array} \right) \mapsto \left( \begin{array}{c} f_1(x_1,x_2,\ldots,x_n) \\ f_2(x_1,x_2,\ldots,x_n) \\ \ldots \\ f_n(x_1,x_2,\ldots,x_n) \end{array} \right)$$
where
$$ f_i \left( x_1, x_2, \ldots, x_n \right) = \sum\limits_{j_1, \ldots, j_r = 1}^{n} f_i^{j_1, j_2,...,j_r} x_{j_1} x_{j_2} ... x_{j_r} $$
Resultant $R_{n|r}(f)$ vanishes on degenerate maps and only on them. To say it another way, resultant is a common divisor of all polynomials, which vanish on degenerate maps. By this very definition, $R_{n|r}(f)$ is an irreducible polynomial in coefficients of $f$. Irreducibility of resultant is a generalization of the well-known fact, that determinant of a generic matrix is irreducible. For illustration, resultant
$$ R_{3|1}(f) = \left| \begin {array}{ccc}
f_{11}&f_{12}&f_{13}\\
\noalign{\medskip}f_{21}&f_{22}&f_{23}\\
\noalign{\medskip}f_{31}&f_{32}&f_{33}
\end {array} \right| = f_{1 1} f_{2 2} f_{3 3}-f_{1 1} f_{2 3} f_{3 2}-f_{2 1} f_{1 2} f_{3 3}+f_{2 1} f_{1 3} f_{3 2}+f_{3 1} f_{1 2} f_{2 3}-f_{3 1} f_{1 3} f_{2 2} $$
does not decompose into smaller factors. However, for non-generic maps -- say, if some coefficients vanish -- such decomposition does take place. For illustration, the resultant
$$ \left| \begin {array}{ccc}
f_{11}&0&0\\
\noalign{\medskip}f_{21}&f_{22}&f_{23}\\
\noalign{\medskip}f_{31}&f_{32}&f_{33}
\end {array} \right| = f_{1 1} \Big( f_{2 2} f_{3 3} - f_{2 3} f_{3 2} \Big) $$
decomposes into smaller factors. The same holds for higher degrees of equations. For further study of resultant's irreducibility properties, see \cite{IRREDUC}.

Another important property of $R_{n|r}$ is that it has a simple interpretation in terms of polynomials' roots. For example, consider a system of two polynomials of arbitrary degree in two variables:
$$ f(x,y) = \prod\limits_{i = 1}^{r} (x - \alpha_i y) $$

$$ g(x,y) = \prod\limits_{i = 1}^{r} (x - \beta_i y) $$
Resultant of this system is given by a simple formula
$$ R_{2|r} \sim \prod\limits_{i, j = 1}^{r} ( \beta_{i} - \alpha_j ) $$
because the above system is solvable, if and only if $\alpha_j = \beta_{i}$ for some $i,j$. In fact, the right hand side is nothing but the product of values of $f(x,y)$ on 2-vectors $(\beta_1, 1) \ldots (\beta_r, 1)$, which are roots of $g(x,y)$:
\be
 R_{2|r} \sim f({\vec \beta}_1) \ldots f({\vec \beta}_r), \ \ \ {\vec \beta}_i = \mbox{ roots of }g(x,y)
\label{Poisson2d}
\ee
Generalization to the case of $n$ variables is straightforward. Resultant of a system
\[ \left\{ \begin{array}{c}
f_1(x_1, \ldots, x_n) = 0 \\
\noalign{\medskip} f_2(x_1, \ldots, x_n) = 0 \\
\noalign{\medskip} \ldots \\
\noalign{\medskip} f_n(x_1, \ldots, x_n) = 0 \\
\end{array} \right. \]
where $f_1, f_2, \ldots, f_n$ are homogeneous polynomials of degree $r$, is given by a simple formula
\be
R_{n|r} \sim f({\vec \beta}_1) \ldots f({\vec \beta}_N), \ \ \ {\vec \beta}_i = \mbox{ common roots of }f_2(x_1, \ldots, x_n), \ldots, f_n(x_1, \ldots, x_n) \label{PoissonNd}
\ee
The number $N$ of common roots of a system of $(n - 1)$ equations of degrees $r_1, r_2, \ldots, r_{n-1}$ is equal to $$N = r_1 r_2 \ldots r_{n-1}$$ In our case, we have all $r_i = r$, therefore the number of common roots is $N = r^{n-1}$. It immediately follows that $R_{n|r}$ is a homogeneous polynomial in coefficients of the first equation of degree $N$. Since (\ref{PoissonNd}) can be analogously written with any $f_i$ in place of $f_1$, resultant $R_{n|r}$ in fact has degree $N$ in coefficients of any particular equation $f_i$. Consequently, it is a homogeneous polynomial of degree
\eq{Rdeg}{\deg \Big( R_{n|r} \Big) = n N = n r^{n - 1}}
in coefficients of all equations. For $r = 1$, we recover the degree of determinant: $\deg R_{n|1} = n$.
For systems of quadratic equations, $\deg R_{n|1} = 2^n n$, for cubic equations $\deg R_{n|1} = 3^n n$, and so on.

Despite being simple, the formula (\ref{PoissonNd}) -- called \emph{Poisson product formula} -- is not very useful for calculation of resultants, because it is rather hard to find common roots of $(n - 1)$ non-linear equations in $n$ variables. The roots are complicated functions of coefficients, which generally can not be expressed through radicals. Fortunately, Poisson product formula can be rewritten in several ways, which do not contain the roots explicitly.
Such reformulations \cite{LOGDET, CONTOUR} are more convenient in practical calculations.

\subsection{Examples of resultants}

\paragraph{Case $r=1$.} The simplest example is a linear map, for certainty in two variables:
$$
f: \left( \begin{array}{c}
x\\
y
\end{array}\right) \to
\left( \begin{array}{c}
f_0x + f_1y\\
g_0x + g_1y
\end{array}\right)
$$
The map $f$ is degenerate exactly when the linear system
$$
\left\{\begin{array}{c}
f_0 x + f_1 y = 0\\
g_0 x + g_1 y = 0
\end{array}\right.
$$
has a non-vanishing solution, i.e. when its determinant vanishes:
$$
\left|\begin{array}{cc}
f_0& f_1\\
g_0& g_1
\end{array}\right|
=0
$$
Therefore, resultant $R_{2|1}$ is nothing but a $2 \times 2$ determinant:
$$ R_{2|1}(f) = \det_{2\times 2} (f) = \left|\begin{array}{cc}
f_0& f_1\\
g_0& g_1
\end{array}\right| $$
In complete analogy, for linear maps in $n$ variables resultant $R_{n|1}$ is just a $n \times n$ determinant:
$$
R_{n|1}(f) = \det_{n\times n} (f) =
\left|
\begin{array}{cccc}
f_{11}&f_{12}&...&f_{1n}\\
f_{21}&f_{22}&...&f_{2n}\\
...&...&...&...\\
f_{n1}&f_{n2}&...&f_{nn}
\end{array}
\right|
$$
\paragraph{Case $n=2, r=2$.} As an example of non-linear maps, let us consider a quadratic map
$$
f: \left(\begin{array}{c}
x\\
y
\end{array}\right)
\to
\left(\begin{array}{c}
f_0x^2 + f_1xy + f_2y^2\\
g_0x^2 + g_1xy + g_2y^2
\end{array}\right)
$$
Degeneracy of $f$ implies existence of non-zero $x$ and $y$ such that
$$
\left\{\begin{array}{c}
f_0x^2 + f_1xy + f_2y^2=0\\
g_0x^2 + g_1xy + g_2y^2=0
\end{array}\right.
$$
Provided that $f_0\ne 0$ and $g_0\ne 0$ the latter system can be written as:
$$
\left\{\begin{array}{c}
f_0(x-\alpha_1y)(x-\alpha_2y)=0\\
g_0(x-\beta_1y)(x-\beta_2y)=0
\end{array}\right.
$$
The system has non-trivial solution if and only if the polynomials have a common root, i.e. when
$$(\alpha_1-\beta_1)(\alpha_1-\beta_2)(\alpha_2-\beta_1)(\alpha_2-\beta_2)=0$$
Since in this case explicit formulas for roots are known, one can express the resultant trough the coefficients of $f$ and $g$ directly. After the expansion, we obtain a polynomial (if we include a multiplier $f_0^2 g_0^2$):
$$R_{2|2}(f) = f_0^2g_0^2(\alpha_1-\beta_1)(\alpha_1-\beta_2)(\alpha_2-\beta_1)(\alpha_2-\beta_2)=
f_0^2(g_0^2\alpha_1^2+g_1\alpha_1+g_2)(g_0^2\alpha_2^2+g_1\alpha_2+g_2)=
$$
$$
=
f_0^2(g_0^2\alpha_1^2\alpha_2^2+g_0g_1\alpha_1\alpha_2(\alpha_1+\alpha_2)+
g_0g_2((\alpha_1+\alpha_2)^2-2\alpha_1\alpha_2)+g_1^2\alpha_1\alpha_2+g_1g_2(\alpha_1+\alpha_2)+g_2^2)=
$$

$$
= g_0^2f_2^2-g_0g_1f_1f_2+g_0g_2(f_1^2-2f_0f_2)+g_1^2f_0f_2-g_1g_2f_0f_1+f_0^2g_2^2
$$
In this way, we can calculate a resultant of two quadratic equations in two unknowns. Note, that it has degree 4 in map's coefficients, in accord with the general formula $n r^{n-1} = 2 \cdot 2^1 = 4$. However, it is hardly possible to use such an approach, when the degree of equations (or, worse than that, the number of variables) is greater than two. One needs another method to go beyond the simplest examples.

\subsection{Sylvester method}
The most classical resultant calculating method, which allows to go beyond the simplest examples, was found by Sylvester \cite{SYLV}. This method is to express resultant as a determinant of some auxiliary matrix. Such expression tautologically exists for linear maps in two variables
\eq{Sylv1}{
f: \left( \begin{array}{c}
x\\
y
\end{array}\right) \to
\left( \begin{array}{c}
f_0x + f_1y\\
g_0x + g_1y
\end{array}\right) \ \ \ \ \ \ R_{2|1}\{f,g\} = \left| \begin {array}{cc}
f_{0}&f_{1}\\
\noalign{\medskip}g_{0}&g_{1}\\
\end {array} \right|
}
and Sylvester has found, that it also exists for quadratic maps in two variables
\eq{Sylv2}{
f: \left(\begin{array}{c}
x\\
y
\end{array}\right)
\to
\left(\begin{array}{c}
f_0x^2 + f_1xy + f_2y^2\\
g_0x^2 + g_1xy + g_2y^2
\end{array}\right) \ \ \ \ \ \  R_{2|2}\{f,g\} = \left| \begin {array}{cccc}
f_{0}&f_{1}&f_{2}&0\\
\noalign{\medskip}0&f_{0}&f_{1}&f_{2}\\
\noalign{\medskip}g_{0}&g_{1}&g_{2}&0\\
\noalign{\medskip}0&g_{0}&g_{1}&g_{2}\\
\end {array} \right|
}
as well as for cubic maps in two variables
\eq{Sylv3}{
f: \left(\begin{array}{c}
x\\
y
\end{array}\right)
\to
\left(\begin{array}{c}
f_0x^3 + f_1x^2y + f_2xy^2 + f_3 y^3\\
g_0x^3 + g_1x^2y + g_2xy^2 + g_3 y^3
\end{array}\right), \ \ \ \ \ \ R_{2|3}\{f,g\} = \left| \begin {array}{cccccc}
f_{0}&f_{1}&f_{2}&f_{3}&0&0\\
\noalign{\medskip}0&f_{0}&f_{1}&f_{2}&f_{3}&0\\
\noalign{\medskip}0&0&f_{0}&f_{1}&f_{2}&f_{3}\\
\noalign{\medskip}g_{0}&g_{1}&g_{2}&g_{3}&0&0\\
\noalign{\medskip}0&g_{0}&g_{1}&g_{2}&g_{3}&0\\
\noalign{\medskip}0&0&g_{0}&g_{1}&g_{2}&g_{3}\\
\end {array} \right|
}
and for all higher degrees in two variables:
\eq{Sylvn}{
f: \left(\begin{array}{c}
x\\
y
\end{array}\right)
\to
\left(\begin{array}{c}
f_0 x^r + f_1 x^{r-1} y + \ldots + f_r y^r \\
g_0 x^r + g_1 x^{r-1} y + \ldots + g_r y^r
\end{array}\right), \ \ \ R_{2|r}(f)=
\left|
\begin{array}{cccccccc}
f_1&f_2&...&f_r&0&0&...&0\\
0&f_1&f_2&...&f_r&0&...&0\\
...
0&0&...&0&f_1&f_2&...&f_r\\
g_0&g_1&...&g_r&0&0&...&0\\
0&g_0&g_1&...&g_r&0&...&0\\
...
0&0&...&0&g_0&g_1&...&g_r
\end{array}
\right|
}
The beauty and simplicity made this formula and corresponding resultant $R_{2|r}$ widely known.
Note, that it has degree $2r$ in map's coefficients, in accord with the general formula $n r^{n-1} = 2 \cdot r^1 = 2r$.

In our days, Sylvester method is the main apparatus used to calculate two-dimensional resultants and
it is the only piece of resultant theory included in undergraduate text books (see for example \cite{KOSTRIK}).
However, its generalization to $n > 2$ is not straightforward. One possible generalization,
representing resultant as \textit{the determinant of Koszul complex}, is the subject of present paper.
This generalization of Sylvester formula is the most widely known one. Other generalizations,
which produce determinantal formulas for $n > 2$, are still the subject of active research (see, for example, \cite{EISENBUD}).

\subsection{Generalization of Sylvester's method: Koszul complex}
To describe Koszul complexes, we need several definitions. The basic object we deal with are homogeneous polynomials. All homogeneous polynomials $F^{i_1 \ldots i_p} x_{i_1} \ldots x_{i_p}$ of degree $p$ in $n$ variables form a linear space with a monomial basis $x_{i_1} \ldots x_{i_p}$. Introduce now $n$ auxiliary Grassmanian variables $\theta$ such, that
$$ \theta_i \theta_j + \theta_j \theta_i = 0$$
The set of homogeneous polynomials of degree $p$ in $x_1, \ldots, x_n$ and of degree $q$ in $\theta_1, \ldots, \theta_n$ is again a linear space with monomial basis $x_{i_1} \ldots x_{i_p}\theta_{j_1} \ldots \theta_{j_q}$. We will use the notation $\Omega(p,q)$ for this linear space. Note that $q \leq n$: if a polynomial has degree in $\theta$-variables bigger than $n$, it vanishes identically, because $\theta$ are anti-commuting variables. Introduction of such anti-commuting variables leads to a new possibility: given a polynomial map
$$x_j \rightarrow f_j(x) = f_j^{i_1i_2...i_r} x_{i_1}x_{i_2}...x_{i_r}$$
one can construct a linear operator
$$\hat d = f_j(x)\frac{\partial}{\partial\theta_j}$$
which is nilpotent:
$$ \hat d\hat d = f_j(x)f_k(x)\frac{\partial}{\partial\theta_j}\frac{\partial}{\partial\theta_k} = f_j(x)f_k(x) \left( \frac{\partial}{\partial\theta_j}\frac{\partial}{\partial\theta_k} + \frac{\partial}{\partial\theta_k}\frac{\partial}{\partial\theta_j} \right) = 0$$ This operator, called \emph{Koszul differential}, acts on the spaces $\Omega(p,q)$ in the following way:
$$\Omega(p,q)\stackrel{\hat d}{\rightarrow}\Omega(p+r,q-1) \stackrel{\hat d}{\rightarrow} \Omega(p+2r,q-2)\rightarrow ...$$
In this way, operator $\hat d$ generates a sequence of linear spaces, called a \emph{Koszul complex}. If we choose some basis in each linear space, operator $\hat d$ will be represented by a sequence of matrices (generally, not a single matrix). The idea -- generalization of Sylvester's idea -- is that resultant of $f$ can be expressed through minors of these matrices. Before discussing the details, we illustrate this idea with simple examples.

\paragraph{The case of $n=2, r=1$.} The simplest example is a linear map in two variables:
$$
f: \left( \begin{array}{c}
x\\
y
\end{array}\right) \to
\left( \begin{array}{c}
f_0x + f_1y\\
g_0x + g_1y
\end{array}\right)
$$
By definition, differential is
$$\hat d = (f_0x + f_1y)\frac{\partial}{\partial\theta_1}+(g_0x+g_1y)\frac{\partial}{\partial\theta_2}$$
It acts on linear spaces
$$\Omega(1,0)\stackrel{\hat d}{\rightarrow}\Omega(0,1)$$
in the following way:
$$\left( \begin{array}{c}\theta_1 \\ \theta_2 \end{array}\right)
\stackrel{\hat d}{\longrightarrow}
\left( \begin{array}{c}f_0x + f_1y\\g_0x+g_1y\end{array}\right)$$
In the above pair of bases, differential is represented by a $2 \times 2$ matrix:
$$
d =
\left\|\begin{array}{cc}
f_0& f_1\\
g_0& g_1
\end{array}\right\|
$$
and one can see that:
$$
R_{2|1} = \det\ d =
\left|\begin{array}{cc}
f_0& f_1\\
g_0& g_1
\end{array}\right|
$$
This example illustrates the definition of Koszul differential. Of course, it is oversimplified, because the matrix of operator $\hat d$ in this particular case coincides with the matrix of map $f$.

\paragraph{The case of $n=2, r=2$.} In this case the map is
$$
f: \left(\begin{array}{c}
x\\
y
\end{array}\right)
\to
\left(\begin{array}{c}
f_0x^2 + f_1xy + f_2y^2\\
g_0x^2 + g_1xy + g_2y^2
\end{array}\right)
$$
and the differential is
$$\hat d=(f_0x^2 + f_1xy + f_2y^2)\frac{\partial}{\partial\theta_1}+(g_0x^2 + g_1xy + g_2y^2)
\frac{\partial}{\partial\theta_2}$$
It acts on linear spaces
$$\Omega(0,1)\stackrel{\hat d}{\rightarrow}\Omega(2,0)$$
in the following way:
$$
\left(\begin{array}{c}\theta_1 x\\ \theta_2 x\\ \theta_1 y \\ \theta_2y \end{array}\right)
\stackrel{\hat d}{\longrightarrow}
\left( \begin{array}{c}(f_0x^2 + f_1xy + f_2y^2)x\\(g_0x^2 + g_1xy + g_2y^2)x\\(f_0x^2 + f_1xy + f_2y^2)x\\(g_0x^2 + g_1xy + g_2y^2)y\end{array}\right)
$$
In the above pair of bases, differential is represented by a $4 \times 4$ matrix:
$$
d=
\left\|\begin{array}{cccc}
f_0&f_1&f_2&0\\
0&f_0&f_1&f_2\\
g_0&g_1&g_2&0\\
0&g_0&g_1&g_2
\end{array}\right\|
$$
As one can see, we have obtained exactly the Sylvester's matrix. In accord with (\ref{Sylv2})
$$R_{2|2} = \det d = f_{0}^2 g_{2}^2 - f_{1} f_{2} g_{0} g_{1} + f_{1}^2 g_{0} g_{2} - 2 f_{0} f_{2} g_{0} g_{2} + f_{0} f_{2} g_{1}^2 - f_{0} f_{1} g_{1} g_{2} + f_{2}^2 g_{0}^2 $$ And this is not an accident: if $f$ has a non-trivial kernel, then the system
$$
\left\{\begin{array}{ccc}
(f_0x^2+f_1xy+f_2y^2)x=0\\
(f_0x^2+f_1xy+f_2y^2)y=0\\
(g_0x^2+g_1xy+g_2y^2)x=0\\
(g_0x^2+g_1xy+g_2y^2)y=0
\end{array}\right.
$$
has a non-vanishing solution and (since this system is linear in
monomials $x^3, x^2y, xy^2, y^3$) its determinant $ \det\ d $ must
vanish. Therefore, if resultant vanishes, then determinant $\det\ d$
must vanish:
$$ R_{2|2} = 0 \Rightarrow \det\ d = 0$$
For this reason
$$ \det\ d =\mu R_{2|2} $$
Since both $\det\ d$ and $R_{2|2}$ are polynomials of degree $4$ in
$f_0, f_1, g_0, g_1$, the coefficient $\mu$ is just a numeric factor. This example illustrates the use of Koszul differential in the case of non-linear maps.

\paragraph{The case of $n=2, r=1$: an example of 3-term complex.} In both of the examples above, differential $\hat d$ was represented by a single square matrix. Now we will consider an example, where non-trivial complex -- a sequence of several matrices -- arises. We return to linear maps in two variables, where $\hat d$ has a form
$$\hat d = (f_0x + f_1y)\frac{\partial}{\partial\theta_1}+(g_0x+g_1y)\frac{\partial}{\partial\theta_2}$$
We will consider its action on another spaces
$$\Omega(0,2) \stackrel{\hat d}{\rightarrow} \Omega(1,1) \stackrel{\hat d}{\rightarrow}\Omega(2,0)$$
In bases
$$(\theta_1\theta_2),\quad(\theta_1x,\;\theta_1y,\;\theta_2x,\;\theta_2y),\quad (x^2, xy, y^2)$$
differential is represented by a pair of rectangular matrices (not a single square matrix this time):
$$ d_1=\left\|\begin{array}{cccc}-g_0&-g_1&f_0&f_1\end{array}\right\|,
\qquad d_2=\left\|\begin{array}{ccc}f_0&f_1&0\\0&f_0&f_1\\g_0&g_1&0\\0&g_0&g_1\end{array}\right\|$$
Resultant $R_{2|1}$ is closely related to these matrices: the system with the matrix $ d_2$
$$
\left\{
\begin{array}{ccc}
(f_0x+f_1y)x = f_0 x^2 + f_1 xy + 0 y^2 = 0\\
(f_0x+f_1y)y = 0 x^2 + f_0 xy + f_1 y^2 = 0\\
(g_0x+g_1y)x = g_0 x^2 + g_1 xy + 0 y^2 = 0\\
(g_0x+g_1y)y = 0 x^2 + g_0 xy + g_1 y^2 = 0\\
\end{array}\right.
$$
which is a linear system in monomials $x^2,\;xy,\;y^2$, has a non-zero solution if resultant $R_{2|1}$ vanishes. In this case, matrix $ d_2$ becomes degenerate:
$$R_{2|1}=0 \ \ \Rightarrow \ \ \mbox{all } 3 \times 3 \mbox{ minors of } \ d_2 \ \mbox{are vanishing } $$
This is only possible, if all minors of $ d_2$ are divisible by $R_{2|1}$. Therefore, $R_{2|1}$ should be a common factor of all these minors. As direct calculation shows
$$
\left|\begin{array}{ccc}f_0&f_1&0\\0&f_0&f_1\\g_0&g_1&0\end{array}\right|=-f_1\left|\begin{array}{cc}f_0&f_1\\g_0&g_1\end{array}\right|,\qquad
\left|\begin{array}{ccc}f_0&f_1&0\\0&f_0&f_1\\0&g_0&g_1\end{array}\right|=f_0\left|\begin{array}{cc}f_0&f_1\\g_0&g_1\end{array}\right|,
$$
$$
\left|\begin{array}{ccc}f_0&f_1&0\\g_0&g_1&0\\0&g_0&g_1\end{array}\right|=g_1\left|\begin{array}{cc}f_0&f_1\\g_0&g_1\end{array}\right|,\qquad
\left|\begin{array}{ccc}0&f_0&f_1\\g_0&g_1&0\\0&g_0&g_1\end{array}\right|=-g_0\left|\begin{array}{cc}f_0&f_1\\g_0&g_1\end{array}\right|
$$
this is indeed true:

$$\mbox{ common factor } = f_0 g_1 - f_1 g_0 = R_{2|1}$$

This example illustrates that, for one and the same polynomial map, there can be many different Koszul complexes,
differing by the choice of linear spaces $\Omega(p,q)$.
Depending on this choice, the complex can become either simpler or more complicated: in the above examples the simpler complex was
$$\Omega(1,0)\stackrel{\hat d}{\rightarrow}\Omega(0,1)$$
and the more complicated complex was
$$\Omega(0,2) \stackrel{\hat d}{\rightarrow} \Omega(1,1) \stackrel{\hat d}{\rightarrow}\Omega(2,0)$$
In other words, the differential is fixed, but we are free to choose the linear spaces.
In practice, one usually chooses a complex which is as simple as possible.

Moreover, this example illustrates the role of the first matrix $d_1$.
Since resultant is the common factor of minors of $d_2$, we can determine resultant from the matrix $d_2$ only.
From this point of view, it seems that $d_1$ does not provide any information about resultant.
This is true, but only in part: $d_1$ provides precise information about the extraneous factors.
As one can see, the other factors of minors $d_2$ are, up to a sign, elements of $d_1$.
Therefore, $R_{2|1}$ can be found by division of minors of $d_2$ on 'conjugate' minors of $d_1$.
This ratio is a particular example of what we call determinant of a complex.
Remarkably, for Koszul complex the ratio is actually a {\it polynomial}
in the coefficients of the map $f$.

\section{Complexes and Determinants}

\subsection{Complex}
A sequence of linear maps between several linear spaces
$$
0\stackrel{d_0}{\longrightarrow}L_1\stackrel{d_1}{\longrightarrow}...\stackrel{d_{p-1}}{\longrightarrow}L_p\stackrel{d_p}{\longrightarrow}0 $$
is called a $p$-term \textit{complex} iff nilpotency condition
$$
\forall i\quad d_{i + 1} \circ d_{i} = 0
$$
is satisfied. Linear spaces $L_{i}$ of dimensions $ l_i = \dim L_{i} $ are called \emph{terms} of the complex.
Therefore, an ordinary linear map is a 2-term complex (a complex, which consists of two linear spaces and a single map between them).
Operators $\hat d_i$ are called differential operators. According to the standard terminology, if a vector is equal to differential of some other vector (i.e, lies in the image of differential ${\hat d}_i$)
$$ v = {\hat d}_i w $$
it is called \emph{exact}. If a vector is annihilated by differential (i.e, lies in the kernel of ${\hat d}_{i+1}$)
$$ {\hat d}_{i+1} v = 0 $$
it is called \emph{closed}. Equivalently, nilpotency condition can be expressed as
$$
\forall i\quad \mbox{ Im } d_i \subset \mbox{ Ker } d_{i+1}
$$
that is, any exact vector is closed. It is very important, that the inverse is, generally speaking, wrong: there can be vectors, which are closed but not exact. In other words, there can be vectors from Ker $d_{i+1}$ which do not belong to Im $d_i$. Such vectors again form a linear space
$$
H_{i} = \mbox{ Ker } d_{i+1} / \mbox{ Im } d_i
$$
of dimension $h_i = \mbox{ dim }\ H_{i}$, called $i$-th \emph{cohomology} space of $d$. Elements of $H_{i}$ are called cohomologies. A complex, which has no cohomologies (i.e, with $h_i = 0$ for all $i$) is called \emph{exact complex}, because all closed elements in such a complex are exact. The notion of cohomology is so important in theory of complexes, that the whole field is known as "homological" algebra.

\begin{figure}[h]
\begin{center} \includegraphics[width=220pt]{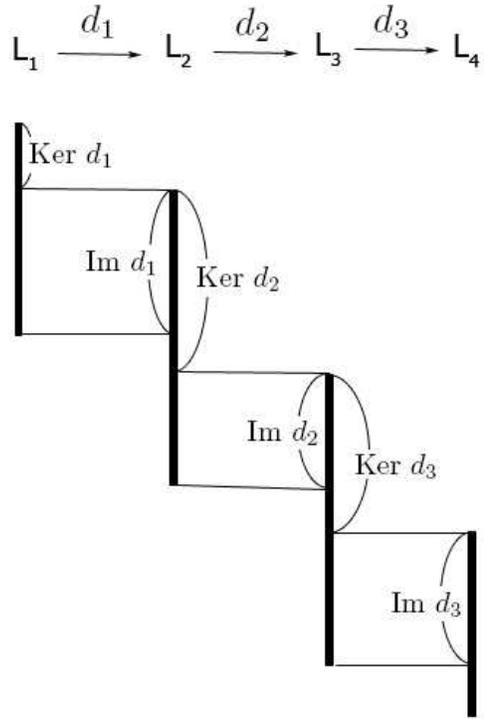}
\caption{A schematic picture of a sequence of linear maps between several linear spaces. }
\smallskip
{It is evident, that $ \mbox{ dim (Im } d_{i}) + \mbox{ dim (Ker } d_{i}) = \mbox{ dim } L_{i} $}
\end{center}
\end{figure}

\pagebreak

It is well known from linear algebra, that
$$ \mbox{ dim (Im } \hat d_{i}) + \mbox{ (dim Ker } \hat d_{i}) = \mbox{ dim } L_{i} $$
At the same time, by definition of cohomology
$$ \mbox{ dim (Im } \hat d_{i}) = \mbox{ dim (Ker } \hat d_{i+1}) - \mbox{ dim } H_{i} $$
From these two identities, it follows
$$ \mbox{ dim } L_{i} = \mbox{ dim (Ker } \hat d_{i}) + \mbox{ dim (Ker } \hat d_{i+1}) + \mbox{ dim } H_{i} $$
If we take an alternating sum, dimensions of the kernel spaces drop away:
\be
\sum\limits_{k = 1}^{p} (-1)^k \mbox{ dim } L_{k} = \sum\limits_{k = 1}^{p} (-1)^k \mbox{ dim } H_{k} \equiv \chi
\label{Euler}
\ee
This alternating combination of dimensions is usually called \emph{Euler characteristic} of the complex and denoted $\chi$. As follows from (\ref{Euler}), if the complex is exact, its Euler characteristic is zero:
\be
\mbox{ complex is exact } \Rightarrow l_1 - l_2 + l_3 - \ldots + (-1)^p l_p = 0
\label{Exactness}
\ee
Inverse is, generally, wrong: vanishing Euler characteristic does not imply all $h_i = 0$,
it implies only that
$$h_1 - h_2 + h_3 - \ldots + (-1)^{p} h_p = 0$$
Consequently, (\ref{Exactness}) is a necessary condition for a complex to be exact.
To say it in other words, if $\chi \neq 0$, the complex is definitely not exact, but if $\chi = 0$,
we need some additional information to decide, whether a given complex is exact or not.
This additional information is provided by determinant of the complex: determinant is non-vanishing only for exact complexes.

\subsection{Determinant of a complex}

In this section we derive an expression for determinant of a complex.
We start from concrete examples, using vector notations and geometric analogies to simplify understanding.
The aim is to create an intuitive image of what determinant of a complex is: a coefficient of proportionality between the two natural geometric quantities. Having this in mind, we will move from simple examples to general cases.

\subsubsection{The case $1 \rightarrow 2 \rightarrow 1$ }

This simplest example is a 3-term complex $
L_1 \stackrel{d_1}{\longrightarrow} L_2 \stackrel{d_2}{\longrightarrow} L_3
$ with dimensions of linear spaces $l = (1,2,1)$. Its Euler characteristic is equal to $1 - 2 + 1 = 0$. The first differential sends
$$ \hat d_1: \ \ x \mapsto {\vec a} x $$
where $x$ is a scalar from $L_1$ and ${\vec a}$ is some constant 2-vector from $L_2$. The second differential sends
$$ \hat d_2: \ \ {\vec y} \mapsto {\vec b} {\vec y} $$
where ${\vec y}$ and ${\vec b}$ are 2-vectors from $L_2$, and ${\vec b} {\vec y} = b_1 y_1 + b_2 y_2$ is a scalar product. As one can see, in this example linear maps $d_1$ and $d_2$ are parametrized by vectors ${\vec a}$ and ${\vec b}$, respectively. Their composition is
$$ \hat d_2 \circ \hat d_1: \ \ x \mapsto {\vec a} {\vec b} \cdot x $$
and the nilpotency condition is that ${\vec a} {\vec b} = 0$, i.e, ${\vec b}$ is orthogonal to ${\vec a}$. In two dimensions, all vectors orthogonal to ${\vec a}$ are proportional to $ \epsilon_{ij} a_{j} $, i.e, to $(a_2, - a_1)$. Vector ${\vec b}$ should be also proportional to it:
$$ b_i = \eta \cdot \epsilon_{ij} a_{j} $$
The coefficient of proportionality $\eta$ is exactly the determinant of this complex. When $\eta \neq 0$, the kernel Ker $\hat d_{2}$ is 1-dimensional and consists of vectors, proportional to ${\vec a}$. It coincides with the image Im $ \hat d_{1}$, so there are no cohomologies and complex is exact. When $\eta$ vanishes, the kernel Ker $ \hat d_{2}$ becomes 2-dimensional and the image Im $\hat d_{1}$ stays 1-dimensional, so 2 - 1 = 1-dimensional cohomologies appear and complex fails to be exact. Thus, $\eta$ indeed allows to distinguish between exact and non-exact complexes.

\begin{figure}[h]
\begin{center} \includegraphics[width=100pt]{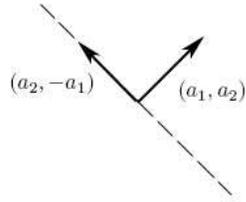}
\caption{Every vector in 2d, orthogonal to $(a_1, a_2)$, is proportional to $(a_2, - a_1)$ }
\end{center}
\end{figure}

Determinant can be written through the components in 2 different ways:
$$ \eta = \dfrac{b_1}{a_2} $$

$$ \eta = - \dfrac{b_2}{a_1} $$
In the third, most symmetric form, determinant is equal to the ratio of lengths
$$ \eta = \dfrac{|b|}{|a|} = \sqrt{ \dfrac{b_1^2 + b_2^2}{a_1^2 + a_2^2} } $$
which emphasizes its orthogonal invariance. Equivalence of these ways follows from $ a_1 b_1 + a_2 b_2 = 0 $.

\subsubsection{The case $2 \rightarrow 3 \rightarrow 1$ }

The next-to-simplest example is a 3-term complex $
L_1 \stackrel{d_1}{\longrightarrow} L_2 \stackrel{d_2}{\longrightarrow} L_3
$ with dimensions of linear spaces $l = (2,3,1)$. Its Euler characteristic is equal to $2 - 3 + 1 = 0$. The first differential sends
$$ \hat d_1: \ \ {\vec x} \mapsto {\vec a} x_1 + {\vec b} x_2 $$
where ${\vec x}$ is a 2-vector from $L_1$ and ${\vec a}, {\vec b}$ is a pair of 3-vectors from $L_{2}$. The second differential sends
$$ \hat d_2: \ \ {\vec y} \mapsto {\vec c} {\vec y}  $$
where ${\vec y}$ and ${\vec c}$ are 3-vectors from $L_2$, and ${\vec c} {\vec y} = c_1 y_1 + c_2 y_2 + c_3 y_3$ is a scalar product. As one can see, in this example linear maps $d_1$ and $d_2$ are parametrized by ${\vec a}, {\vec b}$ and ${\vec c}$, respectively. Their composition is
$$ \hat d_2 \circ \hat d_1: \ \ {\vec x} \mapsto {\vec a} {\vec c} \cdot x_1 + {\vec b} {\vec c} \cdot x_2 $$
with nilpotency condition $${\vec a} {\vec c} = {\vec b} {\vec c} = 0$$ i.e, ${\vec c}$ is orthogonal both to ${\vec a}$ and to ${\vec b}$. In three dimensions, all vectors orthogonal both to ${\vec a}$ and to ${\vec b}$ are proportional to $ \epsilon_{ijk} a_{j} b_{k} $, i.e, to vector product $ {\vec a} \times {\vec b}$. Vector ${\vec c}$ should be also proportional to it:
$${\vec c} = \eta \cdot {\vec a} \times {\vec b} $$

$$ c_i = \eta \cdot \epsilon_{ijk} a_{j} b_{k} $$
The coefficient of proportionality $\eta$ is exactly the determinant of this complex. When $\eta \neq 0$, the kernel Ker $\hat d_{2}$ is 2-dimensional and consists of linear combinations of ${\vec a}$ and ${\vec b}$. It coincides with the image Im $ \hat d_{1}$, so there are no cohomologies and complex is exact. When $\eta$ vanishes, the kernel Ker $ \hat d_{2}$ becomes 3-dimensional and the image Im $\hat d_{1}$ stays 2-dimensional, so 3 - 2 = 1-dimensional cohomologies appear and complex fails to be exact. Thus, $\eta$ indeed allows to distinguish between exact and non-exact complexes.

\begin{figure}[h]
\begin{center} \includegraphics[width=100pt]{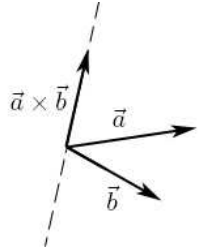}
\caption{Every vector in 3d, orthogonal both to ${\vec a}$ and to ${\vec b}$, is proportional to ${\vec a} \times {\vec b}$ }
\end{center}
\end{figure}

Determinant can be written through the components in 3 different ways:
$$ \eta = \dfrac{c_1}{a_2 b_3 - a_3 b_2} $$

$$ \eta = \dfrac{c_2}{a_3 b_1 - a_1 b_3} $$

$$ \eta = \dfrac{c_3}{a_1 b_2 - a_2 b_1} $$
Equivalence of these three representations follows from
the nilpotency condition
$$ a_1 c_1 + a_2 c_2 + a_3 c_3 = b_1 c_1 + b_2 c_2 + b_3 c_3 = 0 $$

\subsubsection{The case $1 \rightarrow 3 \rightarrow 2$ }

Our next example is a 3-term complex $
L_1 \stackrel{d_1}{\longrightarrow} L_2 \stackrel{d_2}{\longrightarrow} L_3
$ with dimensions of linear spaces $l = (1,3,2)$. Its Euler characteristic is equal to $1 - 3 + 2 = 0$. The first differential sends
$$ \hat d_1: \ \ x \mapsto {\vec a} x $$
where $x$ is a scalar from $L_1$ and ${\vec a}$ is some constant 3-vector from $L_2$. The second differential sends
$$ \hat d_2: \ \ {\vec y} \mapsto \left( \begin{array}{cc} {\vec b} {\vec y} \\ {\vec c} {\vec y} \end{array} \right) $$
where ${\vec y}, {\vec b}$ and ${\vec c}$ are 3-vectors from $L_2$ and the right hand side is a 2-vector from $L_{3}$.
As one can see, in this example linear maps $d_1$ and $d_2$ are parameterized by ${\vec a}$ and ${\vec b}, {\vec c}$, respectively. Their composition is
$$ \hat d_2 \circ \hat d_1: \ \ x \mapsto \left( \begin{array}{cc} {\vec a} {\vec b} \\ {\vec a} {\vec c} \end{array} \right) \cdot x $$
with nilpotency condition $${\vec a} {\vec b} = {\vec a} {\vec c} = 0$$ i.e, ${\vec a}$ is orthogonal both to ${\vec b}$ and to ${\vec c}$.
This implies, in complete analogue with the previous example, that the vector product ${\vec b} \times {\vec c}$ is proportional to ${\vec a}$:
$${\vec b} \times {\vec c} = \eta \cdot {\vec a}$$

$$ \epsilon_{ijk} b_{j} c_{k} = \eta \cdot a_i $$
The coefficient of proportionality $\eta$ is exactly the determinant of this complex. When $\eta \neq 0$, the kernel Ker $\hat d_{2}$ is 1-dimensional and consists of vectors, proportional to ${\vec a}$. It coincides with the image Im $ \hat d_{1}$, so there are no cohomologies and complex is exact. When $\eta$ vanishes, ${\vec b}$ and ${\vec c}$ become collinear and the kernel Ker $ \hat d_{2}$ -- by definition, the space of all vectors, orthogonal to ${\vec b}$ and ${\vec c}$ -- becomes a 2-dimensional plane. The image Im $\hat d_{1}$ stays 1-dimensional, so 2 - 1 = 1-dimensional cohomologies appear and complex fails to be exact. Thus, $\eta$ indeed allows to distinguish between exact and non-exact complexes.

\begin{figure}[h]
\begin{center} \includegraphics[width=200pt]{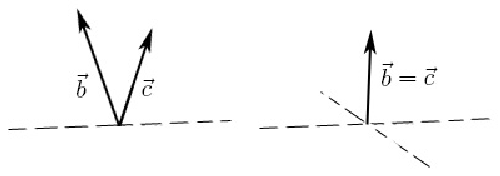}
\caption{The space of vectors, orthogonal to ${\vec b}$ and ${\vec c}$, is 1-dimensional when ${\vec b}$ and ${\vec c}$ are distinct. }
\smallskip
{The space of vectors, orthogonal to ${\vec b}$ and ${\vec c}$, is 2-dimensional when ${\vec b}$ and ${\vec c}$ are collinear.}
\end{center}
\end{figure}

Determinant can be written through the components in 3 different ways:
$$ \eta = \dfrac{b_2 c_3 - b_3 c_2}{a_1} $$

$$ \eta = \dfrac{b_3 c_1 - b_1 c_3}{a_2} $$

$$ \eta = \dfrac{b_1 c_2 - b_2 c_1}{a_3} $$
Again, equivalence of these three representations  follows from
nilpotency conditions
$$ a_1 b_1 + a_2 b_2 + a_3 b_3 = a_1 c_1 + a_2 c_2 + a_3 c_3 = 0 $$

\subsubsection{The case $2 \rightarrow 4 \rightarrow 2$ }

More complicated example is a 3-term complex $
L_1 \stackrel{d_1}{\longrightarrow} L_2 \stackrel{d_2}{\longrightarrow} L_3
$ with dimensions of linear spaces $l = (2,4,2)$. Its Euler characteristic is equal to $2 - 4 + 2 = 0$. The first differential sends
$$ \hat d_1: \ \ {\vec x} \mapsto {\vec a} x_1 + {\vec b} x_2 $$
where $x$ is a 2-vector from $L_1$ and ${\vec a}, {\vec b}$ is a pair of 4-vectors from $L_{2}$. The second differential sends
$$ \hat d_2: \ \ {\vec y} \mapsto \left( \begin{array}{cc} {\vec p} {\vec y} \\ {\vec q} {\vec y} \end{array} \right) $$
where ${\vec y}, {\vec p}$ and ${\vec q}$ are 4-vectors from $L_2$ and the right hand side is a 2-vector from $L_{3}$. As one can see, in this example linear maps $d_1$ and $d_2$ are parametrized by ${\vec a}, {\vec b}$ and ${\vec p}, {\vec q}$, respectively. Their composition is
$$ \hat d_2 \circ \hat d_1: \ \ {\vec x} \mapsto \left( \begin{array}{cc} {\vec p} {\vec a} \\ {\vec q} {\vec a} \end{array} \right) \cdot x_1 + \left( \begin{array}{cc} {\vec p} {\vec b} \\ {\vec q} {\vec b} \end{array} \right) \cdot x_2 $$
with nilpotency condition $${\vec a} {\vec p} = {\vec a} {\vec q} = {\vec b} {\vec p} = {\vec b} {\vec q} = 0$$
Such constraints are quite typical for calculations with 4-vectors. Notice, that
$$ p_{\mu} q_{\nu} - q_{\nu} p_{\mu} $$
is antisymmetric in $\mu, \nu$ and satisfies relations
\[
\begin{array}{ccc}
\big( p_{\mu} q_{\nu} - p_{\nu} q_{\mu} \big) a_{\mu} = 0 \\
\\
\big( p_{\mu} q_{\nu} - p_{\nu} q_{\mu} \big) b_{\mu} = 0 \\
\end{array}
\]
due to nilpotency. Another antisymmetric combination, which satisfies the same relations, is
$$ \epsilon_{\mu \nu \alpha \beta} a_{\alpha} b_{\beta} $$
due to complete antisymmetry of the $\epsilon$-tensor. Consequently, one must be proportional to another:
$$ p_{\mu} q_{\nu} - q_{\nu} p_{\mu} = \eta \cdot \epsilon_{\mu \nu \alpha \beta} a_{\alpha} b_{\beta} $$
The coefficient of proportionality $\eta$ is exactly the determinant of this complex. When $\eta \neq 0$, the kernel Ker $\hat d_{2}$ is 2-dimensional and consists of linear combinations of ${\vec a}$ and ${\vec b}$. It coincides with the image Im $ \hat d_{1}$, so there are no cohomologies and complex is exact. When $\eta$ vanishes, ${\vec p}$ and ${\vec q}$ become collinear and the kernel Ker $ \hat d_{2}$ -- by definition, the space of all vectors, orthogonal to ${\vec p}$ and ${\vec q}$ -- becomes a 3-dimensional (hyper)plane. The image Im $\hat d_{1}$ stays 2-dimensional, so 3 - 2 = 1-dimensional cohomologies appear and complex fails to be exact. Thus, $\eta$ indeed allows to distinguish between exact and non-exact complexes.

\smallskip

Determinant can be written through the components in 6 different ways:
$$ \eta = \dfrac{p_1 q_2 - p_2 q_1}{a_3 b_4 - a_4 b_3} , \ \ \eta = \dfrac{p_1 q_3 - p_3 q_1}{a_2 b_4 - a_4 b_2} $$

$$ \eta = \dfrac{p_1 q_4 - p_4 q_1}{a_2 b_3 - a_3 b_2}, \ \ \eta = \dfrac{p_2 q_3 - p_3 q_2}{a_1 b_4 - a_4 b_1} $$

$$ \eta = \dfrac{p_2 q_4 - p_4 q_2}{a_1 b_3 - a_3 b_1}, \ \ \eta = \dfrac{p_3 q_4 - p_4 q_3}{a_1 b_2 - a_2 b_1} $$
As usual equivalence of all these representations follows from the nilpotency conditions. We do not include a visualization here: a picture of orthogonal vectors in four dimensions is not very illuminating.

\subsubsection{The case $n \rightarrow n + m \rightarrow m$ }

Summarizing the above examples, consider a 3-term complex $
L_1 \stackrel{d_1}{\longrightarrow} L_2 \stackrel{d_2}{\longrightarrow} L_3
$ with dimensions of linear spaces $l = (n, n+m, m)$. Its Euler characteristic is $n - n - m + m = 0$. The first differential sends
$$ {\hat d}_1: \ \ x_i \mapsto A_{ij} x_{j} $$
where $A_{ij}$ is a $n \times (n + m)$ matrix, i.e, index $i$ runs from $1$ to $n$ and index $j$ runs from $1$ to $(n + m)$. \linebreak The second differential sends
$$ {\hat d}_2: \ \ y_i \mapsto B_{ij} y_{j} $$
where $B_{ij}$ is a $(n + m) \times m$ matrix, i.e, index $i$ runs from $1$ to $(n + m)$ and index $j$ runs from $1$ to $m$.
Vector notations, used in the previous examples, are convenient only in low dimensions and we will not use them anymore. From now on, we switch to matrix notations: as one can see, linear maps $d_1$ and $d_2$ are parametrized by matrices $A$ and $B$. Their composition is
$$ {\hat d}_2 \circ {\hat d}_1: \ \ x_i \mapsto A_{ij}B_{jk} \ x_{k} $$
with nilpotency condition $$ A_{ij}B_{jk} = 0 $$
Introduce the minors of matrices $A,B$:
\[
\begin{array}{rl}
\\
\big(M_A\big)_{i_1  \ldots  i_n} = & n \times n \mbox{ minor of A, situated in the coloumns } i_1, \ldots, i_n \\
\\
\big(M_B\big)_{i_1  \ldots  i_n} = & m \times m \mbox{ minor of B, obtained by throwing away the rows } i_1, \ldots, i_n \\
\\
\end{array}
\]
\begin{figure}[t]
\begin{center} \includegraphics[width=300pt]{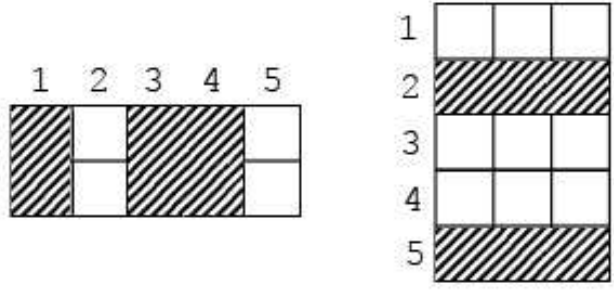}
\caption{Choosing of minors in the case $2 \rightarrow 5 \rightarrow 3$. }
\bigskip
{ Determinant is equal to the ratio of two "white" minors: $\dfrac{\mbox{Minor}_{3 \times 3}}{\mbox{Minor}_{2 \times 2}}$. }
\end{center}
\end{figure}
They are given by explicit expressions with the $\epsilon$-tensors:
\[
\begin{array}{cc}
\big(M_A\big)_{i_1  \ldots  i_n} = \epsilon_{j_1  \ldots  j_n} A_{j_1 i_1} \ldots A_{j_n i_n} \\
\\
\big(M_B\big)_{i_1  \ldots  i_n} = \epsilon_{i_1  \ldots  i_n  j_1  \ldots  j_m} \epsilon_{k_1  \ldots  k_m} B_{j_1  k_1 } \ldots B_{j_m k_m } \\
\end{array}
\]
Notice, that $M_A, M_B$ are both antisymmetric in $i_1, \ldots, i_n$ and vanish when contracted with $B$:
\[
\begin{array}{cc}
\big(M_A\big)_{i_1  \ldots  i_n} \cdot B_{i_1 s_1} = 0 \\
\\
\big(M_B\big)_{i_1  \ldots  i_n} \cdot B_{i_1 s_1} = 0 \\
\end{array}
\]
Tensor $M_A$ does so due to nilpotency conditions, and $M_B$ due to complete antisymmetry of the $\epsilon$-tensor. Consequently, just like it was in the previous example, one must be proportional to another
$$ \big(M_B\big)_{i_1  \ldots  i_n} = \eta \cdot \big(M_A\big)_{i_1  \ldots  i_n} $$
since it is the only way to satisfy such a restrictive set of conditions. In components we have
$$ \epsilon_{i_1  \ldots  i_n  j_1  \ldots  j_m} \epsilon_{k_1  \ldots  k_m} B_{j_1  k_1 } \ldots B_{j_m k_m } = \eta \cdot \epsilon_{j_1  \ldots  j_n} A_{j_1 i_1} \ldots A_{j_n i_n} $$
The coefficient of proportionality $\eta$ is exactly the determinant of this complex. When $\eta \neq 0$, the kernel Ker $\hat d_{2}$ is $n$-dimensional. It coincides with the image Im $ \hat d_{1}$, so there are no cohomologies and complex is exact. When $\eta$ vanishes, all minors of $B$ vanish, so the kernel Ker $\hat d_{2}$ becomes $n + 1$ dimensional. The image Im $\hat d_{1}$ stays $n$-dimensional, so 1-dimensional cohomologies appear and complex fails to be exact. Thus, $\eta$ indeed allows to distinguish between exact and non-exact complexes.

\smallskip

Determinant can be written through components in $C^{n}_{n + m} = (n + m)!/n!/m!$ different ways:
$$ \eta = \mbox{ DET } (A, B) = \dfrac{ \big(M_B\big)_{i_1  \ldots  i_n} }{ \big(M_A\big)_{i_1  \ldots  i_n} }$$
that is the number of components of antisymmetric tensor with $n$ indices running from $1$ to $n+m$. In other words, that is the number of ways to choose $n$ items $i_1, \ldots, i_n$ out of $n + m$ possibilities. Equivalence of these ways, as we have shown, follows from the nilpotency conditions.
On the practical level, different ways correspond to different choices of rows and coloumns (an illustration is shown on Fig. 5).

\begin{figure}[t]
\begin{center} \includegraphics[width=360pt]{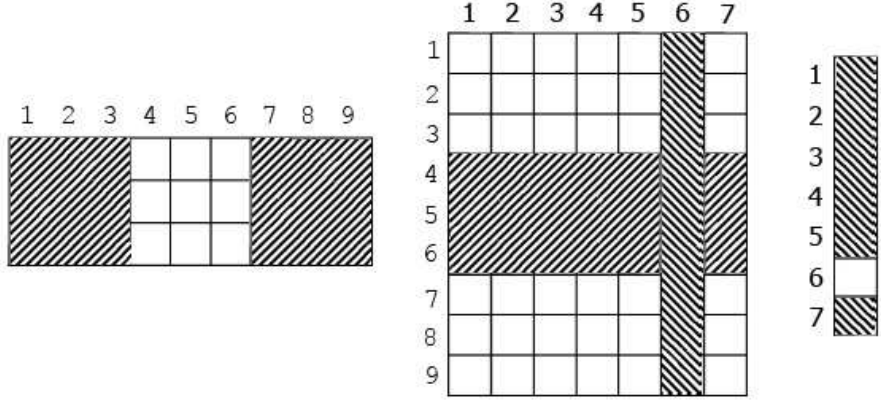}
\caption{Choosing of minors in the case $3 \rightarrow 9 \rightarrow 7 \rightarrow 1$. Determinant is equal to $\dfrac{\mbox{Minor}_{6 \times 6}}{\mbox{Minor}_{3 \times 3}\mbox{Minor}_{1 \times 1}}$ }
\end{center}
\end{figure}

\subsubsection{The case $n \rightarrow n + m \rightarrow m + k \rightarrow k$ }

Moving further, consider a 4-term complex $L_1 \stackrel{d_1}{\longrightarrow} L_2 \stackrel{d_2}{\longrightarrow} L_3 \stackrel{d_3}{\longrightarrow} L_4
$ with dimensions of linear spaces $l = (n, n+m, m+k, k)$. Its Euler characteristic is $n - n - m + m + k - k = 0$. The first differential sends
$$ {\hat d}_1: \ \ x_i \mapsto A_{ij} x_{j} $$
where $A_{ij}$ is a $n \times (n + m)$ matrix, i.e, index $i$ runs from $1$ to $n$ and index $j$ runs from $1$ to $(n + m)$. The second differential sends
$$ {\hat d}_2: \ \ y_i \mapsto B_{ij} y_{j} $$
where $B_{ij}$ is a $(n + m) \times (m + k)$ matrix, i.e, index $i$ runs from $1$ to $(n + m)$ and index $j$ runs from $1$ to $(m + k)$. The third differential sends
$$ {\hat d}_3: \ \ z_i \mapsto C_{ij} z_{j} $$
where $C_{ij}$ is a $(m + k) \times k$ matrix, i.e, index $i$ runs from $1$ to $(m + k)$ and index $j$ runs from $1$ to $k$. Linear maps $d_1, d_2, d_3$ are parametrized by matrices $A, B$ and $C$. Their compositions are equal to
$$ {\hat d}_2 \circ {\hat d}_1: \ \ x_i \mapsto A_{ij}B_{jk} \ x_{k} $$

$$ {\hat d}_3 \circ {\hat d}_2: \ \ y_i \mapsto B_{ij}C_{jk} \ y_{k} $$
with nilpotency conditions
$$ A_{ij}B_{jk} = 0 $$

$$ B_{ij}C_{jk} = 0 $$
The relevant minors of matrices $A,B,C$ are
\[
\begin{array}{lcl}
\\
\big(M_A\big)_{i_1  \ldots  i_n} & = & n \times n \mbox{ minor of A, situated in coloumns } i_1, \ldots, i_n \\
\\
\big(M_B\big)_{i_1  \ldots  i_n \big| j_1 \ldots j_m } & = & m \times m \mbox{ minor of B, obtained by crossing out rows } i_1, \ldots, i_n \\
 & & \mbox{ and selecting coloumns } j_1, \ldots, j_m \\
\\
\big(M_C\big)_{j_1 \ldots j_m } & = & k \times k \mbox{ minor of C, obtained by crossing out rows } j_1, \ldots, j_m \\
\\
\end{array}
\]
They are given by explicit expressions with $\epsilon$-tensors:
\[
\begin{array}{l}
\\
\big(M_A\big)_{i_1  \ldots  i_n} = \epsilon_{j_1  \ldots  j_n} A_{j_1 i_1} \ldots A_{j_n i_n} \\
\\
\big(M_B\big)_{i_1  \ldots  i_n \big| j_1 \ldots j_m } = \epsilon_{i_1 \ldots i_n s_1 \ldots s_m } B_{ s_1 j_1 } \ldots B_{ s_m j_m } \\
\\
\big(M_C\big)_{j_1 \ldots j_m } = \epsilon_{j_1 \ldots j_m s_1 \ldots s_k } \epsilon_{w_1 \ldots w_k} C_{ s_1 w_1  } \ldots C_{ s_k w_k } \\
\\
\end{array}
\]
Let us show, that

$$ \big(M_A\big)_{i_1  \ldots  i_n} \big(M_C\big)_{j_1 \ldots j_m } = \eta \cdot \big(M_B\big)_{i_1  \ldots  i_n \big| j_1 \ldots j_m }$$
\smallskip\\
This is easy: both expressions  vanish when contracted with $B$

$$ \big(M_A\big)_{i_1  \ldots  i_n} \big(M_C\big)_{j_1 \ldots j_m } \cdot B_{i_1 s_1} = \big(M_B\big)_{i_1  \ldots  i_n \big| j_1 \ldots j_m } \cdot B_{i_1 s_1} = 0 $$
\smallskip\\
vanish when contracted with $C$

$$ \big(M_A\big)_{i_1  \ldots  i_n} \big(M_C\big)_{j_1 \ldots j_m } \cdot C_{j_1 s_1} = \big(M_B\big)_{i_1  \ldots  i_n \big| j_1 \ldots j_m } \cdot C_{j_1 s_1} = 0 $$
\smallskip\\
and both are antisymmetric separately in $i_1  \ldots  i_n$ and in $j_1 \ldots j_m$. Consequently, they are proportional, since it is the only way to satisfy such a restrictive set of conditions.

\smallskip

The coefficient of proportionality $\eta$ is exactly the determinant of this complex.
When $\eta \neq 0$, matrices $A$ and $C$ have the full rank, what assures that there is no cohomology. When $\eta$ vanishes, either all minors of $A$ vanish or all minors of $C$ vanish, cohomologies appear either in the first or in the third term, and complex fails to be exact. Thus, $\eta$ indeed allows to distinguish between exact and non-exact complexes.

Determinant can be written through components in $C^{n}_{n + m} C^{m}_{m + k}$ different ways:
$$ \eta = \mbox{ DET } (A, B, C ) = \dfrac{ \big(M_A\big)_{i_1  \ldots  i_n} \big(M_C\big)_{j_1 \ldots j_m } }{ \big(M_B\big)_{i_1  \ldots  i_n \big| j_1 \ldots j_m } }$$
that is the number of ways to choose $n$ items $i_1, \ldots, i_n$ out of $n + m$ possibilities \emph{and} $m$ items $j_1, \ldots, j_m$ out of $m + k$ possibilities. Equivalence of these ways, as we have shown, follows from the nilpotency conditions. On the practical level, different ways correspond to different choices of rows and coloumns (an illustration is shown on Fig. 6).

\subsubsection{The general case}

Finally, consider a $p$-term complex $L_1 \stackrel{d_1}{\longrightarrow} L_2 \stackrel{d_2}{\longrightarrow} \ldots L_{p-1} \stackrel{d_{p-1}}{\longrightarrow} L_p $ with dimensions of linear spaces $$ l_m = k_{m-1} + k_{m} $$ where $k_0$ and $k_{p}$ are by definition zero. Its Euler characteristic is equal to $$ \chi = \sum\limits_{m = 1}^{p} (-1)^{m - 1} (k_{m-1} + k_{m}) = k_{0} + (-1)^{p-1} k_{p} = 0 $$
In fact, it is the general form of a complex with vanishing Euler characteristic. Differentials send
$$ {\hat d}_m: \ \ x_{i} \mapsto \big( d_m \big)_{ij} x_{j} $$
where $\big( d_m \big)_{ij}$ is a $l_m \times l_{m+1}$ matrix, i.e. the index $i$ runs from $1$ to $l_m$ and index $j$
runs from $1$ to $l_{m+1}$. \linebreak In this way our complex is represented by a sequence of rectangular matrices. Compositions are equal to
$$ {\hat d}_{m+1} \circ {\hat d}_m: \ \ x_i \mapsto \big( d_m \big)_{ij}\big( d_{m+1} \big)_{jk} \ x_{k} $$
with nilpotency conditions
$$ \big( d_m \big)_{ij}\big( d_{m+1} \big)_{jk} = 0 $$
\smallskip \\
Just like we did in our previous examples, consider the minors of matrices $d_1, \ldots, d_{p-1}$:
\[
\begin{array}{lcl}
\\
\big(M_1\big)_{\sigma_2} & = & k_1 \times k_1 \mbox{ minor of } d_1 \mbox{, situated in coloumns } \sigma_2 \\
\\
\big(M_2\big)_{\sigma_2 \big| \sigma_3 } & = & k_2 \times k_2 \mbox{ minor of } d_2 \mbox{, obtained by crossing out rows } \sigma_2 \mbox{ and selecting coloumns } \sigma_3 \\
\\
\big(M_3\big)_{\sigma_3 \big| \sigma_4 } & = & k_3 \times k_3 \mbox{ minor of } d_3 \mbox{, obtained by crossing out rows } \sigma_3 \mbox{ and selecting coloumns } \sigma_4 \\
\\
\\
\ldots \ldots \ldots
\\
\\
\big(M_{i}\big)_{\sigma_{i} \big| \sigma_{i+1} } & = & k_{i} \times k_{i} \mbox{ minor of } d_{i} \mbox{, obtained by crossing out rows } \sigma_{i} \mbox{ and selecting coloumns } \sigma_{i+1} \\
\\
\\
\ldots \ldots \ldots
\\
\\
\big(M_{p-1}\big)_{\sigma_{p-1}} & = & k_{p-1} \times k_{p-1} \mbox{ minor of } d_{p-1} \mbox{, obtained by crossing out rows } \sigma_{p-1} \\
\\
\end{array}
\]
where each $ \sigma_i \subset \big\{ 1,2, \ldots, l_i \big\}$ is an arbitrary set of indices of length $\big|\sigma_{i}\big| = k_{i-1}$. \pagebreak They are given by explicit (although a little unwieldy) expressions with $\epsilon$-tensors, just like in the previous examples:
$$ \big(M_1\big)_{ a_1 \ldots a_{ k_1 } } = \epsilon_{s_1 \ldots s_{k_{1}} } \big( d_1 \big)_{ s_1 a_1 } \ldots \big( d_1 \big)_{ s_{k_1} a_{k_1} } $$

$$ \big(M_i\big)_{ a_1 \ldots a_{ k_{i-1} } \big| b_1 \ldots b_{k_i} } = \epsilon_{a_1 \ldots a_{k_{i-1}} s_1 \ldots s_{k_{i}} } \big( d_i \big)_{ s_1 b_1 } \ldots \big( d_i \big)_{ s_{k_{i}} j_{k_{i}} } $$

$$ \big(M_{p-1}\big)_{ a_1 \ldots a_{ k_{p-2} } } = \epsilon_{a_1 \ldots a_{k_{p-2}} s_1 \ldots s_{k_{p-1}} } \epsilon_{j_1 \ldots j_{k_{p-1}}} \big( d_{p-1} \big)_{ s_1 b_1 } \ldots \big( d_{p-1} \big)_{ s_{k_{p-1}} j_{k_{p-1}} } $$
\smallskip\\
Let us show, that
$$ \big(M_1\big)_{\sigma_2} \big(M_3\big)_{\sigma_3 \big| \sigma_4 } \big(M_5\big)_{\sigma_5 \big| \sigma_6 } \ldots = \eta^{\pm 1} \cdot \big(M_2\big)_{\sigma_2 \big| \sigma_3 } \big(M_4\big)_{\sigma_4 \big| \sigma_5 } \big(M_6\big)_{\sigma_6 \big| \sigma_7 } \ldots $$
This is easy: both sides vanish if contracted with matrices $d_i$ (just like in the previous examples,
either due to nilpotency relations or due to complete antisymmetry of $\epsilon$-tensors) and both sides are antisymmetric
under permutations of indices inside sets $\sigma_i$.
These constraints are very restrictive: any two solutions of these constraints must be proportional to each other.

The coefficient of proportionality is either equal to determinant of this complex or equal to inverse determinant of this complex, depending on the number of terms $p$. The easiest way to determine this sign is to notice, that the rightmost differential $d_{p-1}$ should enter the expression for differential in numerator, not in denominator. Consequently,
\be
\eta = \mbox{ DET } (d_1, d_2, \ldots, d_{p-1} ) = \dfrac{\big(M_{p-1}\big)_{\sigma_{p-1}} \big(M_{p-3}\big)_{\sigma_{p-3} \big| \sigma_{p-2} } \big(M_{p-5}\big)_{\sigma_{p-5} \big| \sigma_{p-4} } \ldots }{ \big(M_{p-2}\big)_{\sigma_{p-2} \big| \sigma_{p-1} } \big(M_{p-4}\big)_{\sigma_{p-4} \big| \sigma_{p-3} } \big(M_{p-6}\big)_{\sigma_{p-6} \big| \sigma_{p-5} } \ldots }
\label{DET}
\ee
or, in a slightly more convenient form,
$$ \eta^{(-1)^p} = \dfrac{\big(M_{1}\big)_{\sigma_{2}} \big(M_{3}\big)_{\sigma_{3} \big| \sigma_{4} } \big(M_{5}\big)_{\sigma_{5} \big| \sigma_{6} } \ldots }{ \big(M_{2}\big)_{\sigma_{2} \big| \sigma_{3} } \big(M_{4}\big)_{\sigma_{4} \big| \sigma_{5} } \big(M_{6}\big)_{\sigma_{6} \big| \sigma_{7} } \ldots } $$
By definition of $\sigma_i$, there are $C^{k_1}_{l_2} C^{k_2}_{l_3} \ldots C^{k_{p-2}}_{l_{p-1}}$ different ways to select them. All these choices give one and the same determinant. This independence is ensured by the following facts:
\[
\begin{array}{ll}
1. \mbox{ Numerator and denominator depend on the same indices } \\
\\
2. \mbox{ Numerator and denominator are antisymmetric in each group of indices } \sigma_i \mbox{ separately } \\
\\
3. \mbox{ Numerator and denominator satisfy one and the same highly overdefined system of linear constraints } \\ \mbox{ due to nilpotency of } d \mbox{ and complete antisymmetry of } \epsilon \\
\end{array}
\]

\subsection{Degree of determinant }

As follows from (\ref{DET}) and from the fact, that minor $M_i$ has size $k_i$, resultant has degree
$$ \deg \Big(\mbox{ DET } (d_1, d_2, \ldots, d_{p-1} )\Big) = \sum\limits_{i = 1}^{p-1} (-1)^{p+i+1} k_i = k_{p-1} - k_{p-2} + \ldots $$
in coefficients of the differential operators. In terms of dimensions, we have
$$ k_i = \sum\limits_{j = 1}^{i} (-1)^{i - j} l_{j} = l_{i} - l_{i-1} + \ldots $$

\be
 \deg \Big(\mbox{ DET } (d_1, d_2, \ldots, d_{p-1} )\Big) = \sum\limits_{i = 1}^{p-1} i (-1)^{i-1} l_{p-i} = l_{p-1} - 2 l_{p-2} + \ldots
\label{DegDET}
\ee

\subsection{How to calculate determinant of complex: concise summary }

For reference reasons and also for those, who need to calculate a determinant of their own complex and do not want to spend their time on reading the previous sections, in this section we give a short clear procedure how to calculate a determinant of arbitrary complex
$$
L_1\stackrel{d_1}{\longrightarrow} L_2\stackrel{d_2}{\longrightarrow} ...\stackrel{d_{p-1}}{\longrightarrow}L_p
$$
with vanishing Euler characteristic
$$
\chi = \mbox{dim L}_1 - \mbox{dim L}_2 + \mbox{dim L}_3 - \ldots + (-1)^p \mbox{dim L}_p = 0
$$
Select a basis in each linear space, then complex is represented by a sequence of rectangular matrices. The $l_i \times l_{i+1}$ matrix $d_i$ has $l_i$ rows and $l_{i+1}$ coloums. The determinant of complex $$\eta = \mbox{ DET } (d_1, d_2, \ldots, d_{p-1} )$$ is constructed from minors of these matrices as follows. Take basis vectors in the linear space $L_{i}$ as a set
$$\big\{ 1,2, \ldots, l_i \big\}$$
Select arbitrary subsets $ \sigma_i \subset \big\{ 1,2, \ldots, l_i \big\}$ which consist of
$$ \big| \sigma_i \big| = \sum\limits_{k = 1}^{i-1} (-1)^{i + k + 1} l_k $$
elements, i.e,
$$ \big| \sigma_1 \big| = 0 $$

$$ \big| \sigma_2 \big| = l_1 $$

$$ \big| \sigma_3 \big| = l_2 - l_1 $$

$$ \big| \sigma_4 \big| = l_3 - l_2 + l_1 $$

$$ \big| \sigma_5 \big| = l_4 - l_3 + l_2 - l_1 $$

$$ \ldots \ldots \ldots $$
Define conjugate subsets
$$ \widetilde{ \sigma}_i = \big\{ 1,2, \ldots, l_i \big\} / \sigma_i $$
which consist of
$$\big| \widetilde{ \sigma}_i \big| = l_i - \big| \sigma_i \big| = \big| \sigma_{i+1} \big|$$ elements. Calculate minors
$$ M_i = \mbox{ the minor of } d_i\mbox{, which occupies rows } \widetilde{ \sigma}_i \mbox{ and coloumns } \sigma_{i+1} $$
The determinant of complex equals
\be
\boxed{
\mbox{ DET } (d_1, d_2, \ldots, d_{p-1} ) = \prod\limits_{i = 1}^{p-1} \big( M_{i} \big)^{(-1)^{p + i + 1}} = \dfrac{M_{p-1} M_{p - 3} \ldots }{M_{p - 2} M_{p - 4} \ldots }
}
\ee
In fact, this answer does not depend on the choice of $\sigma_i$. More details are given in the previous section.

\pagebreak

\section{Koszul complexes}

After discussing the general theory of complexes and and their determinants, we return
to Koszul complexes, relevant to the theory of resultants.
The basic objects were already mentioned in the Introduction to the present paper.

Consider a polynomial map of type $n|r$
$$f: \left( \begin{array}{c} x_1 \\ x_2 \\ \ldots \\ x_n \end{array} \right) \mapsto \left( \begin{array}{c} f_1(x_1,x_2,\ldots,x_n) \\ f_2(x_1,x_2,\ldots,x_n) \\ \ldots \\ f_n(x_1,x_2,\ldots,x_n) \end{array} \right)$$
i.e, of degree $r$ in $n$ variables:
$$ f_i \left( x_1, x_2, \ldots, x_n \right) = \sum\limits_{j_1, \ldots, j_r = 1}^{n} f_i^{j_1, j_2,...,j_r} x_{j_1} x_{j_2} ... x_{j_r} $$
Complement the commuting variables
$$ x_1, x_2, \ldots, x_n: \ \ x_i x_j - x_j x_i = 0 $$
with anti-commuting variables
$$ \theta_1, \theta_2, \ldots, \theta_n: \ \ \theta_i \theta_j + \theta_j \theta_i = 0 $$
and consider polynomials, depending both on $x_1, \ldots, x_n$ and $\theta_1, \ldots, \theta_n$.
Denote through $\Omega(p,q) $ the space of such polynomials of degree $p$ in $x$-variables and degree $q$ in $\theta$-variables. Their dimensions are
\be \mbox{ dim } \Omega(p,q) = C^{p}_{p+n-1} C^{q}_{n}
\label{Dims}
\ee
The degree $q$ can not be greater than $n$, since $\theta$ are anti-commuting variables.
{\it Koszul differential}, built from $f$, is a linear operator which acts on these spaces by the rule
$$ \hat d = f_1 (x_1, \ldots, x_n) \dfrac{\partial}{\partial \theta_1} + \ldots + f_n (x_1, \ldots, x_n) \dfrac{\partial}{\partial \theta_n} $$
and is automatically nilpotent:
$$ \hat d\hat d = f_j(x)f_k(x)\frac{\partial}{\partial\theta_j}\frac{\partial}{\partial\theta_k} = f_j(x)f_k(x) \left( \frac{\partial}{\partial\theta_j}\frac{\partial}{\partial\theta_k} + \frac{\partial}{\partial\theta_k}\frac{\partial}{\partial\theta_j} \right) = 0$$
It sends
$$ \hat d: \ \Omega(p,q) \rightarrow \Omega(p + r, q - 1) $$
giving rise to the following {\it Koszul complex}:
$$\Omega(p,q)\stackrel{\hat d}{\rightarrow}\Omega(p+r,q-1)\stackrel{\hat d}{\rightarrow}\Omega(p+2r,q-2)\stackrel{\hat d}{\rightarrow}...\stackrel{\hat d}{\rightarrow}\Omega(0,R)$$
Thus, for one and the same operator $\hat d$ there are many different Koszul complexes,
depending on single integer parameter $R$ -- the $x$-degree of the rightmost space.
Using the formula (\ref{Dims}) for dimensions, it is easy to write down all of them.
For example, the tower of Koszul complexes for the case $3|2$ is
\begin{center}
$\boxed{ 3 \vert 2 }$ \ \
\begin{tabular}{cccc}
R & Spaces & Dimensions & Euler characteristic \\
\hline
$0$ & $\Omega(0,0)$ & $1$ & $1$ \\
$1$ & $\Omega(1,0)$ & $3$ & $3$ \\
$2$ & $\Omega(0,1) \rightarrow \Omega(2,0)$ & $3 \rightarrow 6$ & $3$ \\
$3$ & $\Omega(1,1) \rightarrow \Omega(3,0)$ & $9 \rightarrow 10$ & $1$ \\
$4$ & $\Omega(0,2) \rightarrow \Omega(2,1) \rightarrow \Omega(4,0)$ & $3 \rightarrow 18 \rightarrow 15$ & $0$ \\
$5$ & $\Omega(1,2) \rightarrow \Omega(3,1) \rightarrow \Omega(5,0)$ & $9 \rightarrow 30 \rightarrow 21$ & $0$ \\
$6$ & $\Omega(0,3) \rightarrow \Omega(2,2) \rightarrow \Omega(4,1) \rightarrow \Omega(6,0)$ & $1 \rightarrow 18 \rightarrow 45 \rightarrow 28$ & $0$ \\
$7$ & $\Omega(1,3) \rightarrow \Omega(3,2) \rightarrow \Omega(5,1) \rightarrow \Omega(7,0)$ & $3 \rightarrow 30 \rightarrow 63 \rightarrow 36$ & $0$ \\
$\ldots$ & $\ldots$ & $\ldots$ & $\ldots$
\end{tabular}
\end{center}
For reference reasons, let us write down several other examples explicitly:
\begin{center}
$\boxed{ 3 \vert 3 }$ \ \
\begin{tabular}{cccc}
R & Dimensions & $\chi$ \\
\hline
$0$ & $1$ & $1$ \\
$1$ & $3$ & $3$ \\
$2$ & $6$ & $6$ \\
$3$ & $3 \rightarrow 10$ & $7$ \\
$4$ & $9 \rightarrow 15$ & $6$ \\
$5$ & $18 \rightarrow 21$ & $3$ \\
$6$ & $3 \rightarrow 30 \rightarrow 28$ & $1$ \\
$7$ & $9 \rightarrow 45 \rightarrow 36$ & $0$ \\
$8$ & $18 \rightarrow 63 \rightarrow 45$ & $0$ \\
$9$ & $1 \rightarrow 30 \rightarrow 84 \rightarrow 55$ & $0$ \\
$10$ & $3 \rightarrow 45 \rightarrow 108 \rightarrow 66$ & $0$ \\
$11$ & $6 \rightarrow 63 \rightarrow 135 \rightarrow 78$ & $0$ \\
$\ldots$ & $\ldots$ & $\ldots$
\end{tabular}
\ \ \
$\boxed{ 3 \vert 4 }$ \ \
\begin{tabular}{cccc}
R & Dimensions & $\chi$ \\
\hline
$0$ & $1$ & $1$ \\
$1$ & $3$ & $3$ \\
$2$ & $6$ & $6$ \\
$3$ & $10$ & $10$ \\
$4$ & $3 \rightarrow 15$ & $12$ \\
$5$ & $9 \rightarrow 21$ & $12$ \\
$6$ & $18 \rightarrow 28$ & $10$ \\
$7$ & $30 \rightarrow 36$ & $6$ \\
$8$ & $3 \rightarrow 45 \rightarrow 45$ & $3$ \\
$9$ & $9 \rightarrow 63 \rightarrow 55$ & $1$ \\
$10$ & $18 \rightarrow 84 \rightarrow 66$ & $0$ \\
$11$ & $30 \rightarrow 108 \rightarrow 78$ & $0$ \\
$\ldots$ & $\ldots$ & $\ldots$
\end{tabular}
\end{center}

\begin{center}
$\boxed{ 4 \vert 2 }$ \ \
\begin{tabular}{cccc}
R & Dimensions & $\chi$ \\
\hline
$0$ & $1$ & $1$ \\
$1$ & $4$ & $4$ \\
$2$ & $4 \rightarrow 10$ & $6$ \\
$3$ & $16 \rightarrow 20$ & $4$ \\
$4$ & $6 \rightarrow 40 \rightarrow 35$ & $1$ \\
$5$ & $24 \rightarrow 80 \rightarrow 56$ & $0$ \\
$6$ & $4 \rightarrow 60 \rightarrow 140 \rightarrow 84$ & $0$ \\
$7$ & $16 \rightarrow 120 \rightarrow 224 \rightarrow 120$ & $0$ \\
$\ldots$ & $\ldots$ & $\ldots$
\end{tabular}
\end{center}

\begin{center}
$\boxed{ 4 \vert 3 }$ \ \
\begin{tabular}{cccc}
R & Dimensions & $\chi$ \\
\hline
$0$ & $1$ & $1$ \\
$1$ & $4$ & $4$ \\
$2$ & $10$ & $10$ \\
$3$ & $4 \rightarrow 20$ & $16$ \\
$4$ & $16 \rightarrow 35$ & $19$ \\
$5$ & $40 \rightarrow 56$ & $16$ \\
$6$ & $6 \rightarrow 80 \rightarrow 84$ & $10$ \\
$7$ & $24 \rightarrow 140 \rightarrow 120$ & $4$ \\
$8$ & $60 \rightarrow 224 \rightarrow 165$ & $1$ \\
$9$ & $4 \rightarrow 120 \rightarrow 336 \rightarrow 220$ & $0$ \\
$10$ & $16 \rightarrow 210 \rightarrow 480 \rightarrow 286$ & $0$ \\
$11$ & $40 \rightarrow 336 \rightarrow 660 \rightarrow 364$ & $0$ \\
$12$ & $1 \rightarrow 80 \rightarrow 504 \rightarrow 880 \rightarrow 455$ & $0$ \\
$13$ & $4 \rightarrow 140 \rightarrow 720 \rightarrow 1144 \rightarrow 560$ & $0$ \\
$14$ & $10 \rightarrow 224 \rightarrow 990 \rightarrow 1456 \rightarrow 680$ & $0$ \\
$\ldots$ & $\ldots$ & $\ldots$
\end{tabular}
\ \ $\boxed{ 4 \vert 4 }$ \ \
\begin{tabular}{cccc}
R & Dimensions & $\chi$ \\
\hline
$0$ & $1$ & $1$ \\
$1$ & $4$ & $4$ \\
$2$ & $10$ & $10$ \\
$3$ & $20$ & $20$ \\
$4$ & $4 \rightarrow 35$ & $31$ \\
$5$ & $16 \rightarrow 56$ & $40$ \\
$6$ & $40 \rightarrow 84$ & $44$ \\
$7$ & $80 \rightarrow 120$ & $40$ \\
$8$ & $6 \rightarrow 140 \rightarrow 165$ & $31$ \\
$9$ & $24 \rightarrow 224 \rightarrow 220$ & $20$ \\
$10$ & $60 \rightarrow 336 \rightarrow 286$ & $10$ \\
$11$ & $120 \rightarrow 480 \rightarrow 364$ & $4$ \\
$12$ & $4 \rightarrow 210 \rightarrow 660 \rightarrow 455$ & $1$ \\
$13$ & $16 \rightarrow 336 \rightarrow 880 \rightarrow 560$ & $0$ \\
$14$ & $40 \rightarrow 504 \rightarrow 1144 \rightarrow 680$ & $0$ \\
$\ldots$ & $\ldots$ & $\ldots$
\end{tabular}
\end{center}

\begin{center}
$\boxed{ 5 \vert 2 }$ \ \
\begin{tabular}{cccc}
R & Dimensions & $\chi$ \\
\hline
$0$ & $1$ & $1$ \\
$1$ & $5$ & $5$ \\
$2$ & $5 \rightarrow 15$ & $10$ \\
$3$ & $25 \rightarrow 35$ & $10$ \\
$4$ & $10 \rightarrow 75 \rightarrow 70$ & $5$ \\
$5$ & $50 \rightarrow 175 \rightarrow 126$ & $1$ \\
$6$ & $10 \rightarrow 150 \rightarrow 350 \rightarrow 210$ & $0$ \\
$7$ & $50 \rightarrow 350 \rightarrow 630 \rightarrow 330$ & $0$ \\
$\ldots$ & $\ldots$ & $\ldots$
\end{tabular}
\end{center}
Many properties of Koszul complexes are evident already from these reference tables.
For example, one can see that Euler characteristic $\chi$ vanishes, if $R > n(r-1) $.
If it does, then, according to the previous section, the Koszul complex has a determinant
$$DET \big( d_1, \ldots, d_{p-1} \big)$$
where $ p = 1 + \left[ R/r \right] $ is the number of terms in the complex.
The main fact, announced already in the title of this paper, is that determinant of Koszul complex is equal to the resultant of the original map:
\be
R_{n|r}\big( f_1, \ldots, f_n \big) = DET \big( d_1, \ldots, d_{p-1} \big)
\label{MainEquality}
\ee
Note, that determinant of a complex is generally a rational function; however,
Koszul complexes are very special complexes, and their determinants turn out to be polynomials
(actually, the resultants of the underlying maps $f$).
Despite the importance of this mathematical theorem, we do not include its proof into this text,
it can be found, for example, in the book \cite{GKZ}.
Instead of a proof, we give a number of illustrations and examples,
which do not require any background in homological algebra.

The simplest way to understand the relation (\ref{MainEquality}) is to notice,
that determinant has the same degree in coefficients of $f_1, \ldots, f_n$ as resultant. Indeed,
substituting (\ref{Dims}) into (\ref{DegDET}) we obtain
$$ \deg \Big(\mbox{ DET } (d_1, d_2, \ldots, d_{p-1} )\Big) = \sum\limits_{i = 1}^{p-1} i (-1)^{i-1} l_{p-i} = \sum\limits_{i = 1}^{[R/r]} i (-1)^{i - 1} \mbox{ dim } \Omega( R - ir, i ) = $$

$$ = \sum\limits_{i = 1}^{[R/r]} i (-1)^{i - 1} C^i_{n} C^{R - ir}_{R - ir + n - 1} = \sum\limits_{i = 1}^{[R/r]} \dfrac{ (-1)^{i-1} n }{(n-i)!(i-1)!} \dfrac{(R - ir + n - 1)!}{(R - ir)!} = $$

$$ = \sum\limits_{i = 1}^{[R/r]} \dfrac{ (-1)^{i-1} }{(n-i)!(i-1)!} \oint \dfrac{dz}{z} z^{ir - R} \dfrac{n!}{(1-z)^n} = $$

$$ = \oint \dfrac{dz}{z} \dfrac{n!}{(1-z)^n} \cdot \sum\limits_{i = 1}^{[R/r]} \dfrac{ (-1)^{i-1} }{(n-i)!(i-1)!} z^{ir - R} = $$ $$ = \oint \dfrac{dz}{z} \dfrac{n!}{(1-z)^n} \cdot \dfrac{z^{r - R} (1-z^r)^{n-1}}{(n-1)!} = $$

$$
= n \oint \dfrac{dz (1 - z^r)^{n-1}}{(1-z)^n} z^{r - R - 1} = n \lim\limits_{z \rightarrow 1} \left(\dfrac{1 - z^r}{1 - z}\right)^{n-1} = n r^{n - 1}
$$
\smallskip\\
i.e. if determinant is defined, its degree equals to degree (\ref{Rdeg}) of the resultant:

$$ \deg \Big(\mbox{ DET } (d_1, d_2, \ldots, d_{p-1} )\Big) = \deg \Big( R_{n|r}\big( f_1, \ldots, f_n \big) \Big) $$
\smallskip\\
Consequently,

$$ \mbox{ DET } (d_1, d_2, \ldots, d_{p-1} ) = C (d_1, d_2, \ldots, d_{p-1} ) \cdot R_{n|r}\big( f_1, \ldots, f_n \big) $$
\smallskip\\
where the coefficient of proportionality $ C $ has degree 0.
Since determinant of Koszul complex is polynomial rather than a rational function \cite{GKZ},
$C$ is just a constant and does not depend on anything. This justifies (\ref{MainEquality}).

The second, even more direct, way to establish the relation between resultant and determinant of Koszul complex, is to consider the rightmost differential in the Koszul complex:
$$ \hat d: \ \Omega(R - r, 1) \stackrel{\hat d}{\rightarrow} \Omega(R, 0) $$
which acts on these spaces as

$$ \Big( x_1^{i_1} x_2^{i_2} \ldots x_n^{i_n} \Big) \cdot \theta_j \mapsto \Big( x_1^{i_1} x_2^{i_2} \ldots x_n^{i_n} \Big) \cdot f_j\big( x_1, \ldots, x_n \big) $$
\smallskip\\
and is represented by $n C^{R - r}_{R - r + n - 1} \times C^{R}_{R + n - 1} $ matrix $d_{p - 1}$. If resultant $R_{n|r}\big( f_1, \ldots, f_n \big)$ vanishes, then
\[
\left\{ \begin{array}{c}
f_1(x_1,x_2,\ldots,x_n) = 0 \\
\noalign{\medskip} f_2(x_1,x_2,\ldots,x_n) = 0 \\
\noalign{\medskip} \ldots \\
\noalign{\medskip} f_n(x_1,x_2,\ldots,x_n) = 0 \\
\end{array} \right.
\]
becomes solvable, i.e. a vector $\big( v_1, \ldots, v_n \big)$ exists,
such that all $f_j\big( v_1, \ldots, v_n \big) = 0$.
As a consequence, there is a $n C^{R - r}_{R - r + n - 1}$-dimensional vector
$$ \Big( v_1^{i_1} v_2^{i_2} \ldots v_n^{i_n} \Big) \cdot \theta_j $$
from $\Omega(R - r, 1)$, annihilated by $d_{p - 1}$. Therefore,
$$ R_{n|r} = 0 \Rightarrow \mbox{ rank of }  d_{p - 1} \mbox{ drops down by one } $$
what is the same,

$$ R_{n|r} = 0 \Rightarrow \mbox{ all top-dimensional minors of } d_{p - 1} \mbox{ vanish } $$
\smallskip\\
This is only possible if all top-dimensional minors of $d_{p - 1}$ are divisible by $R_{n|r}$. As follows from (\ref{DET}), determinant of such complex is also divisible by $R_{n|r}$. Whenever resultant $R_{n|r}$ vanishes, determinant of Koszul complex also vanishes. Again, this is enough to state (\ref{MainEquality}), if determinant of Koszul complex is known to be a polynomial.

To finish our discussion of Koszul complexes, let us derive a simple generating function

\be \sum\limits_{R = 0}^{\infty} \chi_R t^R = \left(\dfrac{1 - t^r}{1 - t}\right)^n \label{genfunc} \ee
\smallskip\\
for their Euler characteristics $\chi_R$.
They are quite interesting: for $r = 2$, they coincide with binomial coefficients
(as one can see in the reference tables),
thus for $r > 2$ they provide a generalization of binomial coefficients. Using exactly the same contour integral trick, we obtain

$$ \sum\limits_{R = 0}^{\infty} \chi_R t^R = \sum\limits_{R = 0}^{\infty} \sum\limits_{i = 0}^{\infty} (-1)^{i} \mbox{ dim } \Omega( R - ir, i ) t^R = \sum\limits_{R = 0}^{\infty} \sum\limits_{i = 0}^{\infty} (-1)^{i} C^i_{n} C^{R - ir}_{R - ir + n - 1} t^R = $$

$$ = \sum\limits_{R = 0}^{\infty} \sum\limits_{i = 0}^{\infty} \dfrac{(-1)^i n}{i!(n-i)!} \dfrac{(R - ir + n - 1)!}{(R - ir)!} t^R = \sum\limits_{R = 0}^{\infty} \sum\limits_{i = 0}^{\infty} \dfrac{(-1)^i}{i!(n-i)!} t^R \oint \dfrac{dz}{z} z^{ir - R} \dfrac{n!}{(1-z)^n} = $$

$$ = \oint \dfrac{dz}{z} \dfrac{n!}{(1-z)^n} \cdot \sum\limits_{R = 0}^{\infty} \sum\limits_{i = 0}^{\infty} \dfrac{(-1)^i}{i!(n-i)!} t^R z^{ir - R} = $$ $$ = \oint \dfrac{dz}{z} \dfrac{n!}{(1-z)^n} \cdot \dfrac{z(1-z^r)^n}{n!(z-t)} = $$

$$ = \oint \dfrac{dz}{z - t} \left(\dfrac{1 - z^r}{1 - z}\right)^n = \left(\dfrac{1 - t^r}{1 - t}\right)^n $$
\smallskip\\
and the generating function (\ref{genfunc}) is derived.

\section{Evaluation of $R_{3|2}$ }
As a final illustration to the whole paper, let us consider in detail the case $3|2$, a system of three quadratic equations in three unknowns:
$$
\left\{ \begin{array}{l}
F_1 = a_{11} x_1^2 + a_{12} x_1 x_2 + a_{13} x_1 x_3 + a_{22} x_2^2 + a_{23} x_2 x_3 + a_{33} x_{3}^2 \\
\\
F_2 = b_{11} x_1^2 + b_{12} x_1 x_2 + b_{13} x_1 x_3 + b_{22} x_2^2 + b_{23} x_2 x_3 + b_{33} x_{3}^2 \\
\\
F_3 = c_{11} x_1^2 + c_{12} x_1 x_2 + c_{13} x_1 x_3 + c_{22} x_2^2 + c_{23} x_2 x_3 + c_{33} x_{3}^2 \\
\end{array}\right.
$$
This case is very important in resultant theory, for a number of reasons.
First of all, it is the simplest example of multidimensional ($n > 2$) resultants. At the same time, it is quite a representative example.
$R_{3|2}(F_1, F_2, F_3)$
is a homogeneous polynomial of degree $3\cdot 2^2 = 12$ in $a_{ij}, b_{ij}, c_{ij}$.
If completely expanded, this polynomial consists of 21894 monomials and takes ~30-50 pages of A4 paper,
depending on the font, to write it down.
It is quite amazing, that such a simple system of equations has such an enormously lengthy
consistency (solvability) condition.
Today we know, that these 21894 terms are in fact controlled by various hidden structures
\cite{NOLINAL,ID}.
Explicit knowledge of these structures allows to write down $R_{3|2}$ in a reasonably short form. Koszul complex, which represents $R_{3|2}$ through several matrices of small size, is historically the first example of such a structure.

As usual, Koszul differential is
$$\hat d =
F_1\frac{\partial}{\partial\theta_1}+F_2\frac{\partial}{\partial\theta_2}+F_3\frac{\partial}{\partial\theta_3}
$$
The tower of Koszul complexes is
\begin{center}
\begin{tabular}{cccc}
R & Spaces & Dimensions & Euler characteristic \\
\hline
$0$ & $\Omega(0,0)$ & $1$ & $1$ \\
$1$ & $\Omega(1,0)$ & $3$ & $3$ \\
$2$ & $\Omega(0,1) \rightarrow \Omega(2,0)$ & $3 \rightarrow 6$ & $3$ \\
$3$ & $\Omega(1,1) \rightarrow \Omega(3,0)$ & $9 \rightarrow 10$ & $1$ \\
$4$ & $\Omega(0,2) \rightarrow \Omega(2,1) \rightarrow \Omega(4,0)$ & $3 \rightarrow 18 \rightarrow 15$ & $0$ \\
$5$ & $\Omega(1,2) \rightarrow \Omega(3,1) \rightarrow \Omega(5,0)$ & $9 \rightarrow 30 \rightarrow 21$ & $0$ \\
$6$ & $\Omega(0,3) \rightarrow \Omega(2,2) \rightarrow \Omega(4,1) \rightarrow \Omega(6,0)$ & $1 \rightarrow 18 \rightarrow 45 \rightarrow 28$ & $0$ \\
$7$ & $\Omega(1,3) \rightarrow \Omega(3,2) \rightarrow \Omega(5,1) \rightarrow \Omega(7,0)$ & $3 \rightarrow 30 \rightarrow 63 \rightarrow 36$ & $0$ \\
$\ldots$ & $\ldots$ & $\ldots$ & $\ldots$
\end{tabular}
\end{center}
Let us describe in detail the simplest case $R = 4$ and next-to-the-simplest case $R = 5$.
The complex with $R = 4$, is the first determinant-possessing complex (i.e. has a vanishing Euler characteristics):

$$\Omega(0,2)\rightarrow\Omega(2,1)\rightarrow\Omega(4,0)$$
\smallskip\\
with dimensions $ 3 \rightarrow 18 \rightarrow 15$. If we select a basis in $\Omega(0,2)$ as

$$ \{ \theta_2 \theta_3, \theta_1 \theta_3, \theta_1 \theta_2 \} $$
\smallskip\\
a basis in $\Omega(2,1)$ as

$$ \{ x_{1}^2 \theta_{3}, x_{1}^2 \theta_{2}, x_{1}^2 \theta_{1}, x_{1} x_{2} \theta_{3}, x_{1} x_{2} \theta_{2}, x_{1} x_{2} \theta_{1}, x_{1} x_{3} \theta_{3}, x_{1} x_{3} \theta_{2}, $$
$$ x_{1} x_{3} \theta_{1}, x_{2}^2 \theta_{3}, x_{2}^2 \theta_{2}, x_{2}^2 \theta_{1}, x_{2} x_{3} \theta_{3}, x_{2} x_{3} \theta_{2}, x_{2} x_{3} \theta_{1}, x_{3}^2 \theta_{3}, x_{3}^2 \theta_{2}, x_{3}^2 \theta_{1} \} $$
\smallskip\\
and, finally, a basis in $\Omega(4,0)$ as
$$ \{ x_{1}^4, x_{1}^3 x_{2}, x_{1}^3 x_{3}, x_{1}^2 x_{2}^2, x_{1}^2 x_{2} x_{3}, x_{1}^2 x_{3}^2, x_{1} x_{2}^3, x_{1} x_{2}^2 x_{3}, x_{1} x_{2} x_{3}^2, x_{1} x_{3}^3, x_{2}^4, x_{2}^3 x_{3}, x_{2}^2 x_{3}^2, x_{2} x_{3}^3, x_{3}^4 \} $$
\smallskip\\
then Koszul differential is represented by a pair of matrices: one $3 \times 18$
\begin{center}
\includegraphics[width=350pt]{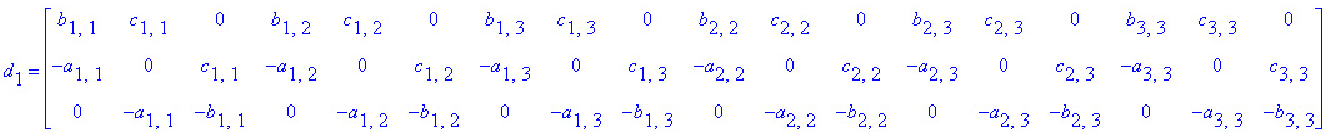}
\end{center}
and another $18 \times 15$
\begin{center}
\includegraphics[width=350pt]{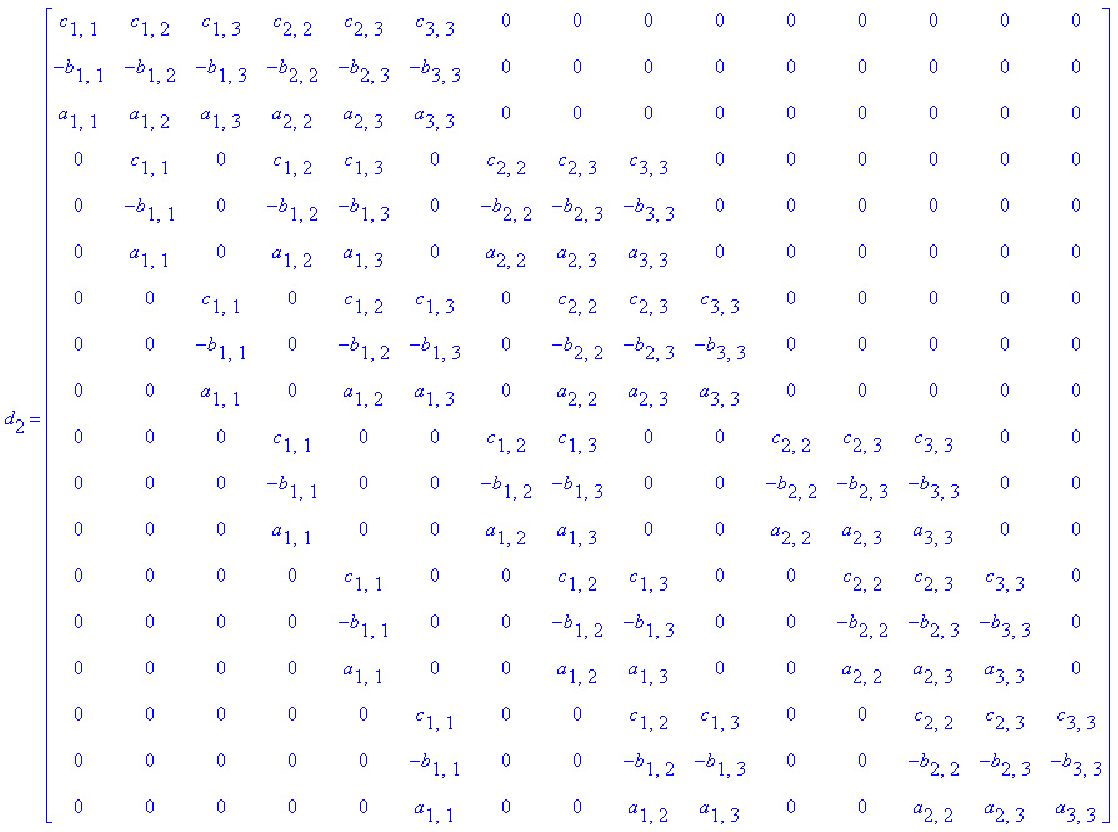}
\end{center}
By selecting some 3 columns in $\hat d_1$ and complementary 15 rows in
$\hat d_2$, we obtain the desired resultant
\begin{center}
\includegraphics[width=300pt]{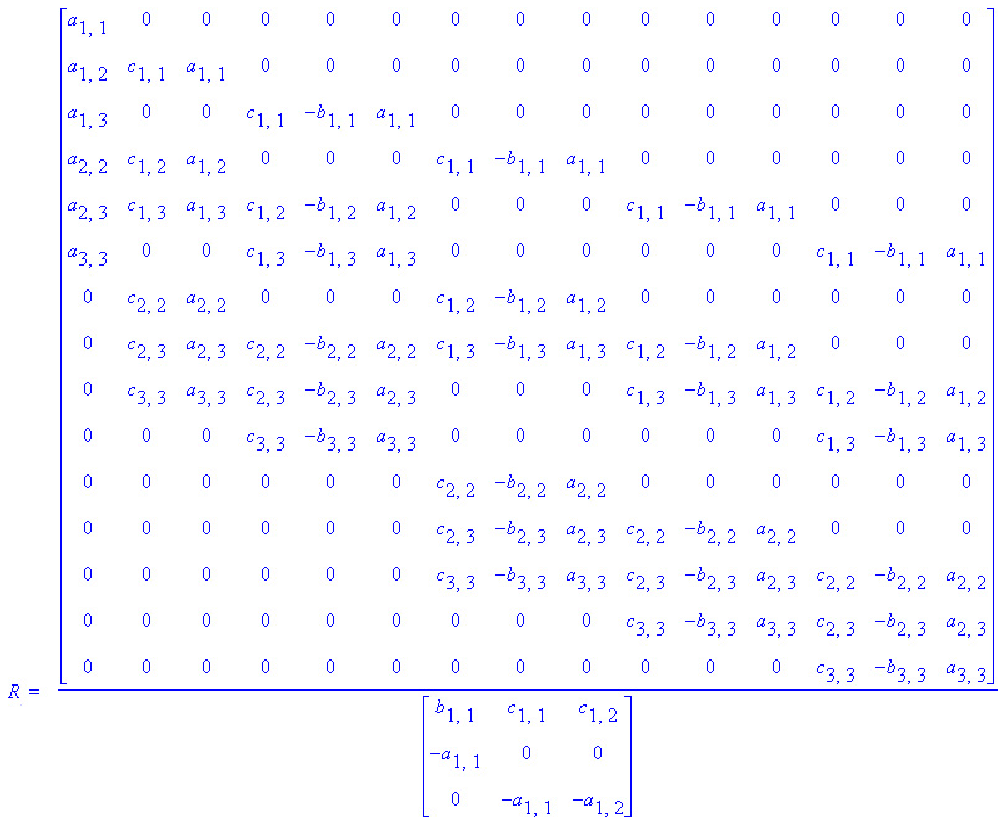}
\end{center}
This particular formula corresponds to the choice of columns 1, 2 and 5
in the first matrix, but the resultant, of course, does not
depend on this choice: any other three columns will do. One only needs
to care that determinant in the denominator does not vanish. If it is
non-zero, then the upper minor is divisible by the lower minor, and
the fraction is equal (up to sign) to the resultant.

Alternative complex is the second complex with vanishing $\chi$, that corresponds to $R = 5$: $$ \Omega(1,2)\rightarrow\Omega(3,1)\rightarrow\Omega(5,0)$$ It has dimensions $ 9 \rightarrow 30 \rightarrow 21$. If we select a basis in $\Omega(1,2)$ as
$$ \{ x_{1} \theta_{2} \theta_{3}, x_{1} \theta_{1} \theta_{3}, x_{1} \theta_{1} \theta_{2}, x_{2} \theta_{2} \theta_{3}, x_{2} \theta_{1} \theta_{3}, x_{2} \theta_{1} \theta_{2}, x_{3} \theta_{2} \theta_{3}, x_{3} \theta_{1} \theta_{3}, x_{3} \theta_{1} \theta_{2} \} $$
a basis in $\Omega(3,1)$ as
$$ \{ x_{1}^3 \theta_{3}, x_{1}^3 \theta_{2}, x_{1}^3 \theta_{1}, x_{1}^2 x_{2} \theta_{3}, x_{1}^2 x_{2} \theta_{2}, x_{1}^2 x_{2} \theta_{1}, x_{1}^2 x_{3} \theta_{3}, x_{1}^2 x_{3} \theta_{2}, x_{1}^2 x_{3} \theta_{1}, $$
$$ x_{1} x_{2}^2 \theta_{3}, x_{1} x_{2}^2 \theta_{2}, x_{1} x_{2}^2 \theta_{1}, x_{1} x_{2} x_{3} \theta_{3}, x_{1} x_{2} x_{3} \theta_{2}, x_{1} x_{2} x_{3} \theta_{1}, x_{1} x_{3}^2 \theta_{3}, x_{1} x_{3}^2 \theta_{2}, x_{1} x_{3}^2 \theta_{1}, x_{2}^3 \theta_{3}, x_{2}^3 \theta_{2}, x_{2}^3 \theta_{1}, $$ $$x_{2}^2 x_{3} \theta_{3}, x_{2}^2 x_{3} \theta_{2}, x_{2}^2 x_{3} \theta_{1}, x_{2} x_{3}^2 \theta_{3}, x_{2} x_{3}^2 \theta_{2}, x_{2} x_{3}^2 \theta_{1}, x_{3}^3 \theta_{3}, x_{3}^3 \theta_{2}, x_{3}^3 \theta_{1} $$
$$ x_{1} x_{3} \theta_{1}, x_{2}^2 \theta_{3}, x_{2}^2 \theta_{2}, x_{2}^2 \theta_{1}, x_{2} x_{3} \theta_{3}, x_{2} x_{3} \theta_{2}, x_{2} x_{3} \theta_{1}, x_{3}^2 \theta_{3}, x_{3}^2 \theta_{2}, x_{3}^2 \theta_{1} \} $$
and, finally, a basis in $\Omega(5,0)$ as
$$ \{ x_{1}^5, x_{1}^4 x_{2}, x_{1}^4 x_{3}, x_{1}^3 x_{2}^2, x_{1}^3 x_{2} x_{3}, x_{1}^3 x_{3}^2, x_{1}^2 x_{2}^3, x_{1}^2 x_{2}^2 x_{3}, x_{1}^2 x_{2} x_{3}^2, x_{1}^2 x_{3}^3, $$ $$ x_{1} x_{2}^4, x_{1} x_{2}^3 x_{3}, x_{1} x_{2}^2 x_{3}^2, x_{1} x_{2} x_{3}^3, x_{1} x_{3}^4, x_{2}^5, x_{2}^4 x_{3}, x_{2}^3 x_{3}^2, x_{2}^2 x_{3}^3, x_{2} x_{3}^4, x_{3}^5 \} $$
then Koszul differential is represented by a pair of matrices: one $9 \times 30$
\begin{center}
\includegraphics[width=400pt]{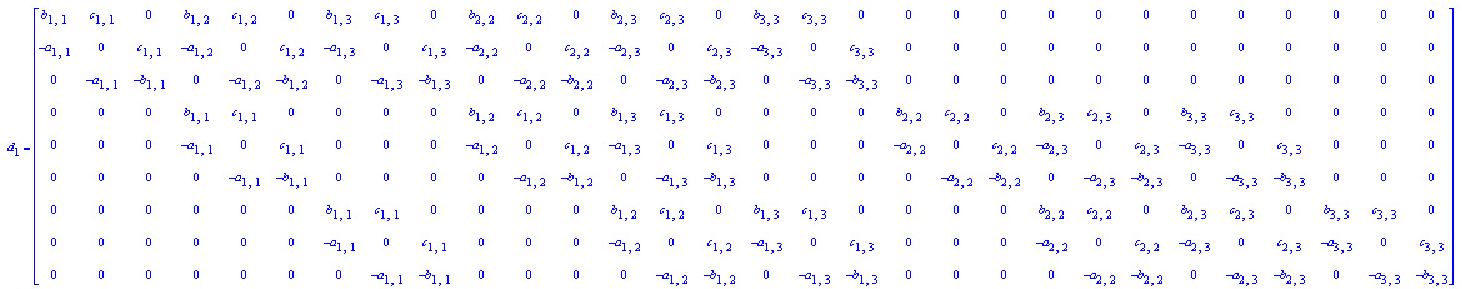}
\end{center}
and another $30 \times 21$
\begin{center}
\includegraphics[width=300pt]{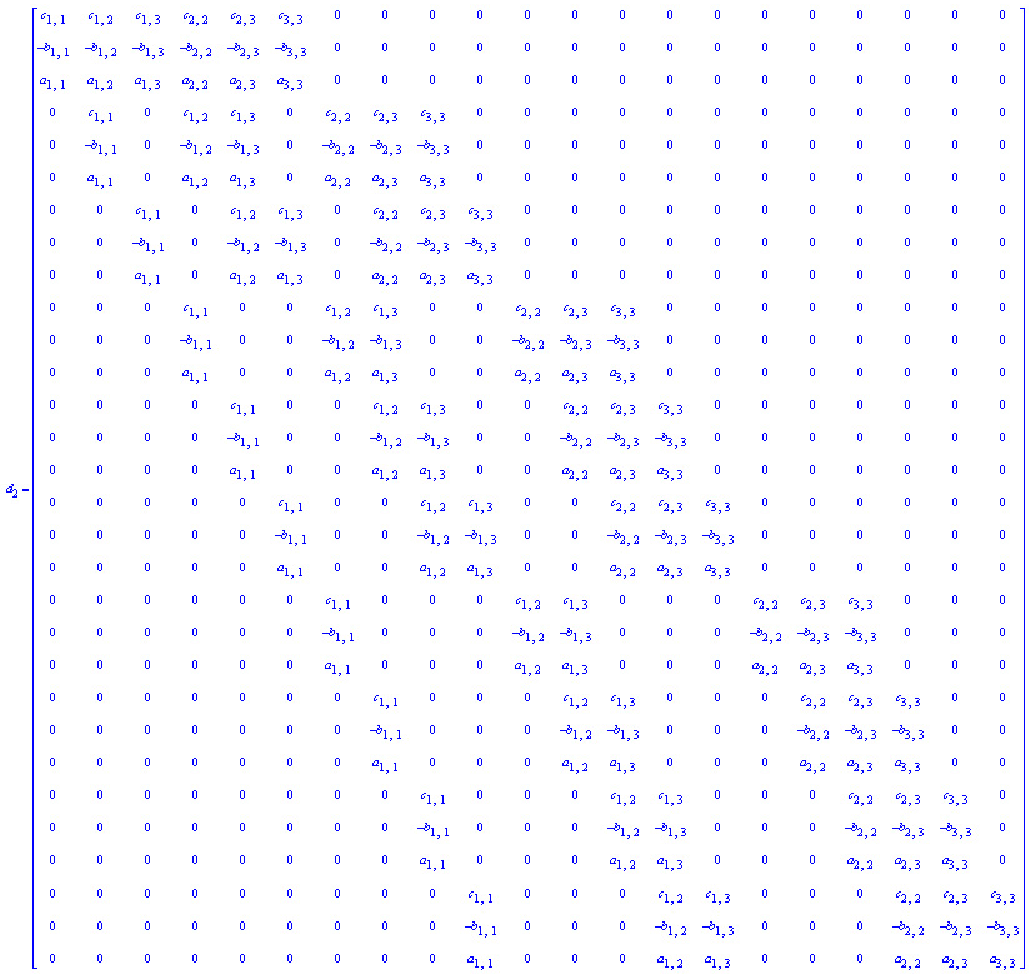}
\end{center}
By selecting some 9 columns in $\hat d_1$ and complementary 21 rows in
$\hat d_2$, we obtain the desired resultant
\begin{center}
\includegraphics[width=450pt]{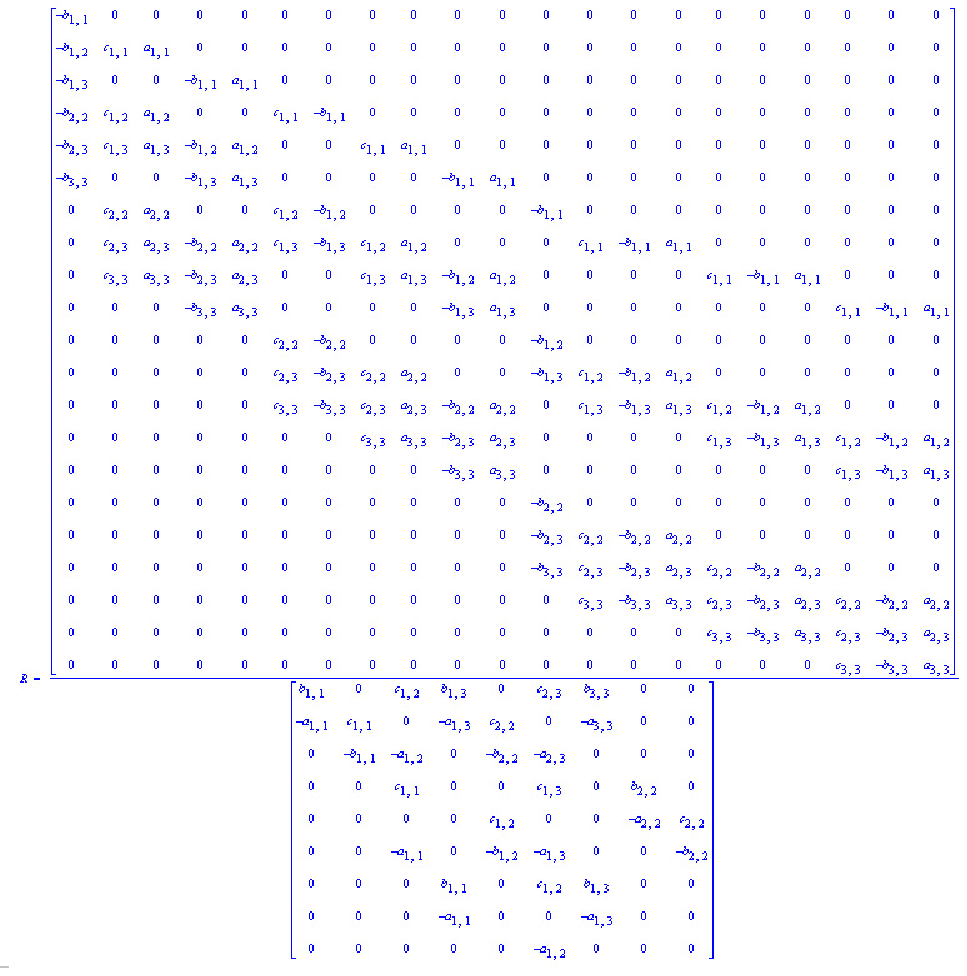}
\end{center}
This particular formula corresponds to the choice of columns
$1,3,5,7,12,14,16,19$ and $21$ in the first matrix. Of
course, any other choice will give the the same (up to a sign) result,
if only the minor in denominator is non-zero. Direct calculation
(division of one minor by another) should demonstrate, that the two complexes $ 3
\rightarrow 18 \rightarrow 15$ and $ 9 \rightarrow 30 \rightarrow
21$ give one and the same expression for the resultant.
It should also be reproduced by determinants of all other complexes
from the $3|2$ list with $R > 3$.

There are alternative representations, providing $R_{3|2}$ in far more
compact form, see, for example, \cite{NOLINAL,LOGDET,ID}.
Still, advantage of Koszul complexes is that they provide an explicit answer
for {\it any} resultant by a straightforward, generic and simply-formulated
algorithm:
\be
{\rm Resultant\ of\ a\ map}\ =\
{\rm Determinant\ of}\ {\it any}\ {\rm associated\ Koszul\ complex}
\label{rede}
\ee
This answer is excessively complicated in the following two senses:
(i) the ratio of various determinants at the r.h.s. is actually reduced to
a polynomial, but this cancelation is not explicit in (\ref{rede});
(ii) the r.h.s. is actually independent of the choice of Koszul complex,
but this independence is not explicit in (\ref{rede}).
However, until powerful alternatives were recently found in
\cite{CONTOUR,RIEM,LOGDET},
relation (\ref{rede}) remained for some years the only
{\it universal} representation for generic resultants.

\section{Conclusion}
This paper is a brief review of three important notions of tomorrow's
mathematical physics: determinant of linear complex,
resultant of non-linear map and Koszul complex.
Each of them has a its own value and large domain of potential
applications. Remarkably, they are interrelated:
resultant is a determinant of a Koszul complex.
We gave a number of simple illustrations to this general
theorem which do not require any preliminary knowledge, but reveal all
essential nuances and  ambiguities.
Still further simplification of this presentation would be important
for making these technical means into a nicely developed and widely
used tool of theoretical physics and string theory.

\section*{Acknowledgements}

We are indebted for illuminating explanations
and discussions to V.Dolotin and A.Gorodentsev.
Our work is partly supported
by Russian Federal Nuclear Energy Agency
and the Russian President's Grant of
Support for the Scientific Schools NSh-3036.2008.2,
by RFBR grants 07-01-00526 (A.A.)
and 07-02-00645 (A.M. \& Sh.Sh.),
by the joint grants 09-01-92440-CE and 09-02-91005-ANF
and by the  NWO project 047.011.2004.026 (A.M.). The work of Sh.Shakirov is also supported in part by the Moebius Contest Foundation for Young Scientists and by the Dynasty Foundation.


\begin{thebibliography}{40}

\bibitem{HISTORY} A. Jennings, \emph{Matrices, ancient and modern}, Bull. Inst. Math. Appl. \textbf{13 (5)} (1977) 117 -- 123

\bibitem{GAUSS} E. Forbes, \emph{The astronomical work of Carl
Friedrich Gauss (1777-1855)}, Historia Math. \textbf{5 (2)} (1978) 167 -- 181

\bibitem{CAYLEY} A. Cayley, \emph{On the theory of elimination}, Cambridge and Dublin Mathematical J. \textbf{3} (1848) 116 - 120 \\
F.S. Macaulay , \emph{On some Formulae in Elimination}, Proceedings of The London Mathematical Society, Vol. \textbf{XXXV} (1903) 3 - 27 \\
A.L. Dixon, \emph{The eliminant of three quantics in two independent
variables}, Proceedings of The London Mathematical Society, \textbf{6} (1908) 468 - 478 \\
E. B\'ezout, \emph{Th\'eorie g\'en\'erale des Equations Alg\'ebriques}, 1779, Paris

\bibitem{SYLV} J.J. Sylvester , \emph{On a general method of determining
by mere inspection the derivations from two equations of any degree}, Philosophical Magazine \textbf{16} (1840) 132 - 135

\bibitem{GKZ} I. Gelfand, M. Kapranov, and A. Zelevinsky,
\emph{Discriminants, Resultants and Multidimensional Determinants}, Birkhauser, 1994

\bibitem{NOLINAL} V. Dolotin and A. Morozov,
{\it Introduction to Non-Linear Algebra}, World Scientific, 2007; arXiv:hep-th/0609022

\bibitem{APP1}
A. Miyake and M. Wadati, \emph{Multiparticle Entaglement and
Hyperdeterminants}, arXiv:quant-ph/02121146;\\
M. Duff, \emph{String Triality, Black Hole Entropy and Cayleyes Hyperdeterminant}, arXiv:hep-th/0601134;
\emph{Hidden Symmetries of the Nambu-Goto Action}, arXiv:hep-th/0602160;\\
A. Linde and R. Kallosh,
\emph{Strings, Black Holes and Quantum Information}, arXiv:hep-th/0602061

\bibitem{APP2}
S. Kachru, A. Klemm, W. Lerche, P. Mayr and C. Vafa,
\emph{Nonperturbative Results on the Point Particle Limit of N=2
Heterotic String Compactifications}, Nucl.Phys. \textbf{B459} (1996) 537-558, arXiv:hep-th/9508155; \\
T. Eguchi and Y. Tachikawa, \emph{Rigid Limit in N=2 Supergravity and Weak-Gravity Conjecture}, arXiv:hep-th/0706.2114; \\
P. Aspinwall, B. Greene and d. Morrison, \emph{Measuring Small
Distances in N=2 Sigma Models}, Nucl.Phys. \textbf{B420} (1994) 184-242,
arXiv:hep-th/9311042

\bibitem{APP3} A. Morozov and A. Niemi, \emph{Can Renormalization Group Flow End in a Big Mess?} Nucl.Phys. \textbf{B666} (2003) 311-336, arXiv:hep-th/0304178; \\
V. Dolotin and A. Morozov, \emph{The Universal Mandelbrot Set.
Beginning of the Story}, arXiv:hep-th/0501235;
Int.J.Mod.Phys. \textbf{A23} (2008) 3613-3684, arXiv:hep-th/0701234;   \\
An. Morozov, \emph{Universal Mandelbrot Set as a Model of Phase Transition Theory},
JETP Lett. \textbf{86} (2007) 745-748, arXiv:nlin/0710.2315

\bibitem{APP4} D. Manocha,
\emph{Algebraic and Numeric Techniques for Modeling and Robotics}, PhD thesis,
Computer Science Division, Department of Electrical Engineering and Computer Science, University of
California, Berkeley.

\bibitem{CONTOUR} A. Morozov and Sh. Shakirov, \emph{Resultants and Contour Integrals}, arXiv:0807.4539

\bibitem{RIEM} B. Gustafsson and V.Tkachev, \emph{The resultant on compact Riemann surfaces}, arXiv:math/0710.2326, to appear in Comm.Math.Phys

\bibitem{TOPOLOG}  E. Witten, \textit{Topological Sigma Models}, Commun. Math. Phys. \textbf{118} (1988) 411;\\
E. Witten, \textit{Mirror Manifolds And Topological Field Theory}, hep-th/9112056;\\
A.Losev, I.Polyubin, {\it On Connection between Topological Landau-Ginzburg Gravity and Integrable Systems}, hep-th/9305079;\\
M. Alexandrov, M. Kontsevich, A. Schwarz and O. Zaboronsky,
\textit{The Geometry of the master equation and topological
quantum field theory}, Int. J. Mod. Phys. A \textbf{12} (1997) 1405; hep-th/9502010;\\
A. Losev, N. Nekrasov, S. Shatashvili,
{\it Issues in Topological Gauge Theory}, hep-th/9711108

\bibitem{BV}
L.D. Faddeev and V.N. Popov, Phys. Lett. \textbf{B 25} (1967) 29; \\
I. V. Tyutin, \textit{Gauge Invariance In Field Theory And Statistical Physics In Operator Formalism}, LEBEDEV-75-39;\\
R.Stora, \textit{Progress in Gauge Field Theory} New York: Plenum
Press, 1984;\\
I. Batalin and G. Vilkoviski, Nucl. Phys. \textbf{B234} (1984) 106;\\
A. Schwarz, \textit{Geometry of Batalin-Vilkovisky quantization}, Commun. Math. Phys. \textbf{155} (1993) 249; hep-th/9205088

\bibitem{bra}
M. Douglas, \textit{D-branes, Categories and N=1 Supersymmetry},
J.Math.Phys. \textbf{42} (2001) 2818-2843, arXiv:hep-th/0011017;\\
G. Moore, \textit{Some Comments on Branes, G-flux, and K-theory},
Int.J.Mod.Phys. \textbf{A16} (2001) 936-944; \\
C.I. Lazaroiu, \textit{Unitarity, D-brane dynamics and D-brane categories},
JHEP \textbf{0112} (2001) 031, arXiv:hep-th/0102183; \\
D.E. Diaconescu, \textit{Enhanced D-Brane Categories from String Field Theory},
JHEP {\textbf 0106} (2001) 016, arXiv:hep-th/0104200; \\
A. Kapustin and D. Orlov, \textit{Remarks on A-branes, Mirror Symmetry, and the Fukaya category},
J.Geom.Phys. {\textbf 48} (2003) 84, arXiv:hep-th/0109098; \\ J. Distler, H. Jockers and H. Park,
\textit{D-Brane Monodromies, Derived Categories and Boundary Linear
Sigma Models}, arXiv:hep-th/0206242; \\
Sh. Katz and E. Sharpe, \textit{D-branes, open string vertex operators, and Ext groups},
Adv.Theor.Math.Phys. \textbf{6} (2003) 979-1030,
arXiv:hep-th/0208104; \\
P.S. Aspinwall, \textit{D-Branes on Calabi-Yau Manifolds},  hep-th/0403166;\\
O. Lechtenfeld, A.D. Popov and R.J. Szabo, \textit{Rank Two Quiver
Gauge Theory, Graded Connections and Noncommutative Vortices},
JHEP \textbf{0609} (2006) 054, arXiv:hep-th/0603232; \\
A. Kapustin and E. Witten, \textit{Electric-Magnetic Duality And The Geometric Langlands Program},
arXiv:hep-th/0604151

\bibitem{LOGDET} A. Morozov and Sh. Shakirov, \emph{Analogue of the identity Log Det = Trace Log for resultants}, arXiv:math-ph/0804.4632

\bibitem{EISENBUD} D. Eisenbud and F.O. Schreyer, \emph{Resultants and Chow forms via exterior syzygies}, J. Amer. Math. Soc. \textbf{16} (2003) 537-579, arXiv:math/0111040

\bibitem{ALGO}
C. Andrea, A.Dickenstein, \emph{Explicit formulas for the multivariate resultant}, arXiv:math/0007036; \\
M. Chardin. \emph{Formules \`a la Macaulay pour les sous-resultants en plusieurs variables}, C. R. Acad. Sci. Paris, \textbf{319} (1994) 433 -- 436; \\
d. Manocha and J.Canny, \emph{Multipolynomial Resultant Algorithms}, J.Symb.Comp. \textbf{15} (1993) 99 -- 122; \\
J. Canny, E.Kaltofen and L.Yagati, \emph{Solving Systems of Non-Linear Equations Faster}, Proc.Internat.Symp.Symbolic Algebraic Comput., (1989) 121-128

\bibitem{IRREDUC} A. Ostrowski. \emph{The irreducibility of the resultant and connected irreducibility theorems}. \textbf{29 }(1977) 252-260 \\
L. Buse and C. D'Andrea, \emph{On the irreducibility of multivariate subresultants}, Compt.Rend.Math. \textbf{338(4)} (2004) 287-290, arXiv:math/0309374

\bibitem{KOSTRIK} A. Kostrikin, \emph{Introduction to algebra}, Springer-Verlag, New York, 1982

\bibitem{ID}
A. Morozov and Sh. Shakirov, \emph{Introduction to Integral Discriminants}, to appear

\end{thebibliography}
\end{document}